\documentclass[prd,aps,twocolumn,showpacs,preprintnumbers,amsmath,amssymb,nofootinbib]{revtex4-1}

\RequirePackage[T1]{fontenc}
\usepackage[english]{babel}

\usepackage{newtxtext}
\usepackage{newtxmath}

\usepackage{booktabs}
\usepackage{bm}
\usepackage{caption} 
\usepackage{graphicx}
\usepackage{xstring}
\usepackage{braket}
\RequirePackage{flushend}
\RequirePackage{xcolor}

\newcommand\Eprint[2]{%
  \StrBefore{#2}{ [}[\EprintClean]%
  \IfStrEq{\EprintClean}{}{\href{#1}{#2}}{\href{#1}{\EprintClean}}%
}

\RequirePackage{hyperref}
\hypersetup{
    linktocpage,
    colorlinks,
    citecolor=blue,
    filecolor=black,
    linkcolor=blue,
    urlcolor=blue,
}

\begin{document}

\title{Is there gluonium in the $\eta^\prime$? The flavour structure of $\eta$ and $\eta^\prime$ mixing in radiative $V\to P\gamma$ and $P\to V\gamma$ decays}

\author{Rafel Escribano$^{1,2}$}
\email{rescriba@ifae.es}
\author{Emilio Royo$^{2,3}$}
\email{emilio.royocarratala@uchceu.es}

\affiliation{
$^1$Grup de F\'{i}sica Te\`{o}rica, Departament de F\'{i}sica, Universitat Aut\`{o}noma de Barcelona, 08193 Bellaterra (Barcelona), Spain\\
$^2$Institut de F\'{i}sica d'Altes Energies (IFAE) and The Barcelona Institute of Science and Technology (BIST), Campus UAB, 08193 Bellaterra (Barcelona), Spain\\
$^3$Departamento de Matem\'{a}ticas, F\'{i}sica y Ciencias Tecnol\'{o}gicas, Universidad Cardenal Herrera-CEU, CEU Universities, 46115 Alfara del Patriarca, Val\`{e}ncia, Spain
}


\begin{abstract}

Radiative $V\to P\gamma$ and $P\to V\gamma$ decays are conventionally described using constituent-quark mass differences and phenomenological wavefunction-overlap factors as sources of flavour-symmetry breaking. We investigate whether the overlap factors can instead be understood as the effect of mixing of the $\eta$ and $\eta^\prime$ with an additional pseudoscalar state, described by an $SO(4)$ rotation among four basis states. A global fit to current data yields a significantly better description than the corresponding three-state limit without overlap factors, while preserving the established $\eta$--$\eta^\prime$ and vector-meson mixing angles, and shows that the additional state mixes predominantly with the non-strange component. The conventional overlap-factor description is recovered exactly as the special case in which this mixing is purely non-strange. The non-strange pattern disfavours both a flavour-singlet admixture, as expected from mixing with a pseudoscalar glueball, and the gluonium scenario considered in previous analyses. Radiative decays therefore provide no compelling evidence for gluonium in the $\eta^\prime$, and the overlap factors are better interpreted in terms of a radially excited non-strange state.

\end{abstract}



\maketitle


\section{Introduction}
\label{sec:intro}

The flavour-singlet pseudoscalar sector occupies a unique position in low-energy QCD. Unlike the octet of Goldstone bosons, whose properties are largely governed by spontaneous chiral symmetry breaking and explicit $SU(3)_F$ symmetry breaking, the singlet state is directly affected by the $U(1)_A$ axial anomaly~\cite{Adler:1969gk,Bell:1969ts,Fujikawa:1979ay}. Through the Witten--Veneziano mechanism~\cite{Witten:1979vv,Veneziano:1979ec,Cichy:2015jra}, the anomaly is responsible for the large $\eta^\prime$ mass, making the $\eta$--$\eta^\prime$ system a particularly sensitive probe of non-perturbative QCD dynamics~\cite{Feldmann:1999uf}. 

Because the flavour-singlet channel couples directly to gluons through the anomaly, the $\eta^\prime$ may contain a sizeable gluonic component. This possibility has been debated for several decades~\cite{Bass:2018xmz,Gan:2020aco} and has been invoked, for instance, to explain the unexpectedly large $\eta^\prime$ production observed in $B$-meson decays~\cite{Ball:1995zv,Fritzsch:1997ps}. It has recently attracted renewed interest with the identification of the $\eta(2370)$, formerly $X(2370)$, as a glueball-dominated state by the BESIII Collaboration~\cite{BESIII:2023wfi,BESIII:2026mvn}, and with a lattice-QCD study of the mixing between the pseudoscalar glueball and the flavour-singlet pseudoscalar meson, which finds this mixing to be small~\cite{Jiang:2022ffl}.

Radiative transitions between light vector and pseudoscalar mesons provide one of the cleanest phenomenological laboratories for studying pseudoscalar mixing~\cite{Feldmann:1998vh,Feldmann:1998sh,Escribano:1999nh,Escribano:2005qq}. Within the framework of Vector Meson Dominance (VMD)~\cite{Sakurai:1960ju,Kroll:1967it,Lurie:1968zz,Bando:1985rf,Bando:1987br,Meissner:1987ge}, the $SU(3)_F$-symmetric effective Lagrangian relates the measured decay widths directly to the pseudoscalar and vector mixing angles. Building on this framework, Bramon \textit{et al.}~\cite{Bramon:2000fr} introduced two independent mechanisms to account for $SU(3)_F$ symmetry breaking. The first originates from differences in the effective magnetic moments of light and strange quarks in magnetic dipole transitions and is incorporated via constituent-quark mass differences.\footnote{Specifically, this is implemented by means of a multiplicative $SU(3)_F$-breaking term, i.e.~$1-s_e\equiv\bar{m}/m_s$, in the $s$-quark entry of the quark-charge matrix $Q$; see Eq.~(\ref{eq:Q}).} The second is associated with differences in the spatial extensions of the meson wavefunctions and is parametrised by additional multiplicative factors accounting for the corresponding relative wavefunction overlaps~\cite{Bramon:2000fr}.

This phenomenological framework has also been used to search for gluonic components in the $\eta$ and $\eta^\prime$ wavefunctions, with conflicting conclusions. Early analyses of radiative $V\to P\gamma$ decays suggested a sizeable gluonic admixture in the $\eta^\prime$, with values reaching up to $Z_{\eta^\prime}^2\sim0.26$~\cite{Kou:1999tt}. A dedicated measurement of $\phi\to\eta^\prime\gamma$ by the KLOE Collaboration later reported $Z_{\eta^\prime}^2=0.14\pm0.04$~\cite{KLOE:2006guu}, superseding an earlier, less precise determination consistent with zero~\cite{KLOE:2002jed}, and a subsequent global fit gave $Z_{\eta^\prime}^2=0.12\pm0.04$~\cite{Ambrosino:2009sc}. In contrast, Ref.~\cite{Escribano:2007cd} found no significant evidence for such a component once the overlap factors were included, a conclusion also reached in Ref.~\cite{Thomas:2007uy}. The origin of this ambiguity is that a gluonic admixture and an overlap factor both suppress the quark--antiquark contributions to the $\eta$ and $\eta^\prime$ couplings, so that the two effects are difficult to disentangle.

The experimental situation has evolved considerably since then. The radiative decay widths of the light vector and pseudoscalar mesons are now known with substantially improved precision~\cite{ParticleDataGroup:2026aaa}, allowing these processes to serve as stringent tests of pseudoscalar mixing. In particular, Ref.~\cite{Escribano:2020jdy} revisited the phenomenological framework of Bramon \textit{et al.} by incorporating first-order isospin-breaking corrections. The analysis showed that the original implementation of the model was no longer sufficient to describe the current data. A significantly improved description was obtained by extending the second flavour-symmetry-breaking mechanism---namely the wavefunction-overlap factors---to include isospin-breaking effects.

The success of this refined phenomenological description naturally raises a further question. Are the overlap factors simply effective parameters accounting for flavour-dependent symmetry-breaking effects, or are they also absorbing dynamics that are not explicitly included in the pseudoscalar mixing scheme? Motivated by this question, in the present work we explore an alternative interpretation of the current data.

To this end, we retain the constituent-quark mass mechanism of $SU(3)_F$ breaking and replace the phenomenological wavefunction-overlap factors by an explicit extension of the pseudoscalar mixing sector. Specifically, we introduce a single additional pseudoscalar degree of freedom and formulate the most general mixing among the four orthogonal basis states $\{\pi_3,\eta_{\rm NS},\eta_{\rm S},X_0\}$ by means of an $SO(4)$ rotation matrix. The additional state $X_0$ is not identified \textit{a priori} with any particular physical resonance. Rather, it serves as a generic placeholder for dynamics not included in the conventional three-state mixing scheme, which could originate from gluonic degrees of freedom~\cite{Cheng:2008ss,Mathieu:2009sg}, from higher pseudoscalar excitations~\cite{Cohen:1978ge,Klempt:2007cp}, or from other flavour-singlet contributions~\cite{Damgaard:1993sx}. The fit determines not only the size of the admixture of $X_0$ in the physical states but also its flavour structure. This provides a direct test of the gluonium interpretation, since a gluonic admixture induced by the flavour-blind anomaly is expected to follow the flavour-singlet direction, whereas other mechanisms need not.

The paper is organised as follows. Section~\ref{sec:model} introduces the extended pseudoscalar mixing framework and derives the radiative decay couplings, while Sec.~\ref{sec:fit} presents the global fits to the data and discusses the results. Section~\ref{sec:interpretation} explores possible physical interpretations of the solution favoured by the data, and Sec.~\ref{sec:conclusions} presents our conclusions.


\section{The Extended Pseudoscalar Mixing Framework}
\label{sec:model}

The phenomenological framework developed in this work extends the conventional treatment of radiative vector--pseudoscalar transitions by enlarging the pseudoscalar sector, while retaining the standard VMD structure of the interaction Lagrangian. The central hypothesis is that the wavefunction-overlap factors, which constitute the secondary source of flavour-symmetry breaking in the conventional scheme~\cite{Bramon:2000fr,Escribano:2020jdy}, can be replaced by an explicit extension of this sector. This section introduces the extended mixing scheme and derives the corresponding radiative decay couplings.

Radiative transitions between light vector and pseudoscalar mesons are described within the VMD framework by the effective Lagrangian~\cite{Bramon:1997va}
\begin{equation}
\mathcal{L}_{V\!P\gamma}
=
g\,
\epsilon_{\mu\nu\alpha\beta}\,
\partial^\mu A^\nu\,
\mathrm{Tr}
\!\left[
Q
\left(
\partial^\alpha V^\beta P
+
P\,\partial^\alpha V^\beta
\right)
\right] ,
\label{eq:Lag}
\end{equation}
where $A^\mu$ denotes the electromagnetic field, $V^\mu$ and $P$ are the vector and pseudoscalar meson matrices, respectively, and $g$ is an effective electromagnetic coupling. The charge matrix
\begin{equation}
Q=
\mathrm{diag}
\left(
\frac23,
-\frac13,
-\frac13\,\frac{\bar{m}}{m_s}
\right)
\label{eq:Q}
\end{equation}
implements the leading $SU(3)_F$ breaking through the constituent-quark mass ratio $\bar{m}/m_s$ (with $\bar{m}$ the isospin-averaged mass of the light quarks and $m_s$ the mass of the strange quark), which accounts for the reduced magnetic moment of the strange quark~\cite{Bramon:2000fr}.

The pseudoscalar meson matrix is written in the quark-flavour basis as
\[
P=
\begin{pmatrix}
\dfrac{\pi_3+\eta_{\rm NS}}{\sqrt2}
&
\pi^+
&
K^+
\\[3mm]
\pi^-
&
\dfrac{-\pi_3+\eta_{\rm NS}}{\sqrt2}
&
K^0
\\[3mm]
K^-
&
\bar K^0
&
\eta_{\rm S}
\end{pmatrix}\, ,
\]
where $\pi_3$ denotes the third component of the isospin triplet, and the non-strange and strange states are given by
\[
|\eta_{\rm NS}\rangle=\frac{1}{\sqrt{2}}(|u\bar u\rangle+|d\bar d\rangle)\, ,\qquad
|\eta_{\rm S}\rangle=|s\bar s\rangle\, .
\]

The vector meson matrix is given analogously in the quark-flavour basis,
\[
V=
\begin{pmatrix}
\dfrac{\rho^0+\omega_{\rm NS}}{\sqrt2}
&
\rho^+
&
K^{*+}
\\[3mm]
\rho^-
&
\dfrac{-\rho^0+\omega_{\rm NS}}{\sqrt2}
&
K^{*0}
\\[3mm]
K^{*-}
&
\bar K^{*0}
&
\omega_{\rm S}
\end{pmatrix}\, ,
\]
where the non-strange and strange states $\omega_{\rm NS}$ and $\omega_{\rm S}$ have the same quark content as $\eta_{\rm NS}$ and $\eta_{\rm S}$.
The physical states $\omega$ and $\phi$ are expressed as mixtures of these basis states,
\[
\begin{aligned}
\omega&=\cos\phi_V\,\omega_{\rm NS}-\sin\phi_V\,\omega_{\rm S}\, ,\\
\phi&=\sin\phi_V\,\omega_{\rm NS}+\cos\phi_V\,\omega_{\rm S}\, ,
\end{aligned}
\]
where $\phi_V$ parametrises the deviation from ideal vector mixing.

The extension proposed in this work concerns the pseudoscalar sector. We enlarge the conventional basis with an additional isoscalar pseudoscalar state $X_0$, orthogonal to the ground-state nonet, without making any assumption about its microscopic nature or its flavour structure; it is the minimal extension of the pseudoscalar sector required to test the central hypothesis of this work. The physical pseudoscalar states are then related to the orthonormal basis states by an $SO(4)$ rotation $R$,
\begin{equation}
\begin{pmatrix}\pi^0\\ \eta\\ \eta^\prime\\ X\end{pmatrix}
=
R\,
\begin{pmatrix}\pi_3\\ \eta_{\rm NS}\\ \eta_{\rm S}\\ X_0\end{pmatrix}\, ,
\label{eq:states}
\end{equation}
which we parametrise as the product of elementary rotations $\mathcal{R}_{ij}$ by an angle $\phi_{ij}$ in the $(i,j)$ plane,
\begin{equation}
R=\mathcal{R}_{23}\,\mathcal{R}_{12}^{T}\,\mathcal{R}_{13}^{T}\,\mathcal{R}_{14}^{T}\,\mathcal{R}_{24}^{T}\,\mathcal{R}_{34}^{T}\, ,
\label{eq:Rdef}
\end{equation}
where, for instance,
\[
\mathcal{R}_{23}=
\begin{pmatrix}
1&0&0&0\\
0&\cos\phi_{23}&-\sin\phi_{23}&0\\
0&\sin\phi_{23}&\cos\phi_{23}&0\\
0&0&0&1
\end{pmatrix}\, .
\]
The angles $\phi_{12}$ and $\phi_{13}$ describe the isospin-breaking admixtures of $\eta_{\rm NS}$ and $\eta_{\rm S}$ in the $\pi^0$. Since they are small, we treat them to first order, $\sin\phi_{1j}\simeq\epsilon_{1j}$ $(j=2,3)$, as in Ref.~\cite{Escribano:2020jdy}, while all remaining angles are retained exactly. The explicit form of $R$ is given in Appendix~\ref{app:matrix}.

In the three-state limit $(\phi_{14}=\phi_{24}=\phi_{34}=0)$, the rotation reduces to that of Ref.~\cite{Escribano:2020jdy}, and $\phi_{23}$ plays the role of the usual $\eta$--$\eta^\prime$ mixing angle in the quark-flavour basis. Neglecting isospin breaking,
\begin{equation}
\begin{aligned}
\eta&=\cos\phi_{23}\,\eta_{\rm NS}-\sin\phi_{23}\,\eta_{\rm S}\, ,\\
\eta^\prime&=\sin\phi_{23}\,\eta_{\rm NS}+\cos\phi_{23}\,\eta_{\rm S}\, ,
\end{aligned}
\label{eq:etamixing}
\end{equation}
and $\phi_{23}$ is related to the octet--singlet mixing angle $\theta_P$ by $\phi_{23}=\theta_P+\arctan\sqrt2$. This relation no longer holds once $X_0$ mixes with the $\eta$ and $\eta^\prime$.\footnote{If the rotation is parametrised in the same way in the octet--singlet basis $\{\pi_3,\eta_8,\eta_0,X_0\}$, with angles $\theta_{ij}$, the angle $\theta_{23}$ generalises $\theta_P$ and the relation becomes $\phi_{23}=\theta_{23}+\delta$, with $\tan\delta=\sqrt2\cos\theta_{34}/(\cos\theta_{24}-\sqrt2\sin\theta_{24}\sin\theta_{34})$ up to isospin-breaking terms of order $\epsilon_{1j}\sin\theta_{14}$. For $\theta_{24}=\theta_{34}=0$ this gives $\delta=\arctan\sqrt2\simeq54.7^\circ$, while for the extended fit of Sec.~\ref{sec:fit} $\delta\simeq56.4^\circ$.}

The three new angles $\phi_{14}$, $\phi_{24}$ and $\phi_{34}$ describe the mixing with $X_0$. Their meaning is most easily understood from the fourth physical state which, using the shorthand $c_{ij}\equiv\cos\phi_{ij}$ and $s_{ij}\equiv\sin\phi_{ij}$, reads
\begin{equation}
X=-s_{14}\,\pi_3-c_{14}\left(s_{24}\,\eta_{\rm NS}+c_{24}s_{34}\,\eta_{\rm S}\right)+c_{14}c_{24}c_{34}\,X_0\, .
\label{eq:X}
\end{equation}
This shows that $\phi_{14}$ controls the isospin-breaking mixing of $X_0$ with the pion, while the term in brackets, governed by $\phi_{24}$ and $\phi_{34}$, is the isoscalar quark--antiquark part of the admixture. Its flavour content is conveniently characterised by a single angle $\alpha$, the direction of this term in the $(\eta_{\rm NS},\eta_{\rm S})$ plane,
\begin{equation}
\tan\alpha=\frac{c_{24}s_{34}}{s_{24}}\, .
\label{eq:alpha}
\end{equation}
A purely non-strange admixture corresponds to $\alpha=0$, a purely strange one to $\alpha=90^\circ$, and a flavour-singlet admixture, along $\eta_0=(\sqrt2\,\eta_{\rm NS}+\eta_{\rm S})/\sqrt3$, to $\alpha=\arctan(1/\sqrt2)\simeq35.3^\circ$. Unlike the individual angles, which depend on the choice of basis and on the ordering of the rotations in Eq.~(\ref{eq:Rdef}), $\alpha$ and the composition of the physical states are basis independent, and our conclusions on the flavour structure of the mixing are based on them.

\subsection{Radiative Decay Couplings}

The effective Lagrangian of Eq.~(\ref{eq:Lag}) contains no explicit coupling to the additional basis state $X_0$. Its effect on the radiative decay couplings therefore arises exclusively through its admixture into the $\pi^0$, $\eta$ and $\eta^\prime$. Since these states are normalised, an $X_0$ component reduces the weight of their quark--antiquark components and hence their couplings.

The radiative decay couplings are obtained by evaluating the flavour trace in Eq.~(\ref{eq:Lag}) and expressing the basis states in terms of the physical fields. To this end, we write each physical pseudoscalar $P\in\{\pi^0,\eta,\eta^\prime\}$ in terms of the basis states,
\begin{equation}
P=a_P^{\pi}\,\pi_3+a_P^{\rm NS}\,\eta_{\rm NS}+a_P^{\rm S}\,\eta_{\rm S}+a_P^{X}\,X_0\, ,
\label{eq:components}
\end{equation}
where the amplitudes of the $\pi^0$, $\eta$ and $\eta^\prime$ are the elements of the first, second and third rows of $R$, respectively (see Appendix~\ref{app:matrix}). Using the shorthand $c_V\equiv\cos\phi_V$ and $s_V\equiv\sin\phi_V$, the couplings of the neutral vector mesons read
\begin{equation}
\begin{aligned}
g_{\rho^0P\gamma}&=g\left(\tfrac13\,a_P^{\pi}+a_P^{\rm NS}\right) ,\\
g_{\omega P\gamma}&=g\left[c_V\left(a_P^{\pi}+\tfrac13\,a_P^{\rm NS}\right)+\tfrac23\tfrac{\bar m}{m_s}\,s_V\,a_P^{\rm S}\right] ,\\
g_{\phi P\gamma}&=g\left[s_V\left(a_P^{\pi}+\tfrac13\,a_P^{\rm NS}\right)-\tfrac23\tfrac{\bar m}{m_s}\,c_V\,a_P^{\rm S}\right] .
\end{aligned}
\label{eq:couplings}
\end{equation}
The numerical weights multiplying the amplitudes follow from the quark charges, and the strange amplitudes carry the factor $\bar m/m_s$ of Eq.~(\ref{eq:Q}). The amplitudes $a_P^{X}$ do not appear, because $X_0$ does not couple at the $V\!P\gamma$ vertex. The charged and strange couplings are not affected by the pseudoscalar mixing,
\begin{equation}
\begin{aligned}
g_{\rho^+\pi^+\gamma}&=\frac{g}{3}\, ,\\
g_{K^{*0}K^0\gamma}&=-\frac{g}{3}\left(1+\frac{\bar m}{m_s}\right) ,\\
g_{K^{*+}K^+\gamma}&=\frac{g}{3}\left(2-\frac{\bar m}{m_s}\right) .
\end{aligned}
\label{eq:charged}
\end{equation}
The couplings thus depend on the nine parameters of the model, $\{g,\epsilon_{12},\epsilon_{13},\phi_{14},\phi_{23},\phi_{24},\phi_{34},\phi_V,\bar m/m_s\}$, namely the overall coupling, the pseudoscalar mixing parameters, the vector mixing angle, and the constituent-quark mass ratio. Their explicit expressions are collected in Appendix~\ref{app:couplings}.


\section{Global Fits to Radiative Decay Data}
\label{sec:fit}

The extended $SO(4)$ pseudoscalar mixing framework is tested through global least-squares fits to the available radiative $V\!P\gamma$ decay data~\cite{ParticleDataGroup:2026aaa}. The aim is to assess whether $X_0$ can account for these data without the phenomenological wavefunction-overlap factors employed in the conventional three-state framework and, if so, to determine the flavour structure of the required mixing.

Our primary analysis is based on the ten $V\to P\gamma$ and $P\to V\gamma$ decay channels that do not involve kaons.\footnote{The two-photon decays included in some previous analyses~\cite{KLOE:2006guu,Ambrosino:2009sc} are not considered here, since in the present framework they would require additional assumptions on the two-photon coupling of $X_0$.} The $K^*\to K\gamma$ modes are treated separately because the corresponding decay widths are derived largely from older Primakoff-effect measurements, whose systematic uncertainties are more difficult to assess than those of the remaining channels. Fits including these channels are also presented for comparison.

The experimental decay couplings are obtained from the measured partial widths using
\[
\Gamma(V\to P\gamma)=\frac{|\mathbf p_\gamma|^3}{12\pi}\,g_{V\!P\gamma}^2\, ,\quad
\Gamma(P\to V\gamma)=\frac{|\mathbf p_\gamma|^3}{4\pi}\,g_{V\!P\gamma}^2\, ,
\]
where $|\mathbf p_\gamma|$ is the photon momentum in the rest frame of the decaying particle. The model parameters are determined by minimising the global least-squares function
\begin{equation}
\chi^2
=
\sum_{i}
\frac{
\left(
g_i^{\rm exp}
-
g_i^{\rm th}
\right)^2
}
{\sigma_i^2}\, ,
\label{eq:chi2}
\end{equation}
where $g_i^{\rm th}$ are the couplings of Eqs.~(\ref{eq:couplings}) and~(\ref{eq:charged}), and $\sigma_i$ the uncertainties of the experimental couplings, which include the propagated errors of the measured branching fractions, total decay widths and particle masses.\footnote{All minima are located with global minimisation algorithms within the region $0<\bar m/m_s<1$; outside it, the $\chi^2$ function develops spurious minima. Inside this region the global minimum is unique up to a simultaneous change of sign of $\phi_{14}$, $\phi_{24}$ and $\phi_{34}$. This sign is unphysical, since it corresponds to $X_0\to-X_0$ and $X_0$ does not couple at the $V\!P\gamma$ vertex, and we fix it by the convention $\phi_{24}>0$. A parametrisation of the rotation in the octet--singlet basis yields identical results.}

\begin{table}[t]
\centering
\captionsetup{font=small, skip=2pt}
\caption{Parameters of the extended $SO(4)$ fit to the ten channels without kaons ($\chi^2_{\rm min}/{\rm d.o.f.}=2.1/1$). The last column gives the $1\sigma$ profile-likelihood interval of each angle, obtained with $\Delta\chi^2=1$ and all other parameters refitted.}
\label{tab:fit}
\setlength{\tabcolsep}{6pt}
\renewcommand{\aboverulesep}{0.5pt}
\renewcommand{\belowrulesep}{0.5pt}
\begin{tabular}{lcc}
\toprule
Parameter & Best-fit value & Profile interval \\
\midrule
$g$ & $(0.71 \pm 0.02)\,\text{GeV}^{-1}$ & \\[1pt]
$\epsilon_{12}$ & $(3.6 \pm 1.1)\, \%$ & \\[1pt]
$\epsilon_{13}$ & $(1.9 \pm 1.3)\, \%$ & \\[1pt]
$\phi_{14}$ & $(4.9 \pm 3.5)^\circ$ & $[1.3,\,8.1]^\circ$ \\[1pt]
$\phi_{23}$ & $(39.9 \pm 1.3)^\circ$ & \\[1pt]
$\phi_{24}$ & $(27.5 \pm 3.5)^\circ$ & $[23.7,\,30.8]^\circ$ \\[1pt]
$\phi_{34}$ & $(4.4 \pm 3.9)^\circ$ & $[0.8,\,8.7]^\circ$ \\[1pt]
$\phi_V$ & $(3.7 \pm 0.5)^\circ$ & \\[1pt]
$\bar{m}/m_s$ & $0.64 \pm 0.02$ & \\
\bottomrule
\end{tabular}
\end{table}

\subsection{Reference Three-State Fit}

As a benchmark, we first consider the three-state limit of the present framework, obtained by setting $\phi_{14}=\phi_{24}=\phi_{34}=0$. The additional basis state is then completely decoupled from the conventional pseudoscalar sector, which corresponds to the formalism of Ref.~\cite{Escribano:2020jdy} with the overlap factors fixed to unity. This provides the reference against which the impact of the extended mixing scheme can be assessed.

For the complete data set, including the two $K^*\to K\gamma$ channels, the fit yields $\chi^2_{\rm min}/{\rm d.o.f.}=36.7/6\simeq6.1$, while excluding the kaon channels it gives $\chi^2_{\rm min}/{\rm d.o.f.}=26.6/4\simeq6.7$. As shown in Table~\ref{tab:widths}, the largest tensions arise in the $\omega\to\pi^0\gamma$, $\eta^\prime\to\rho^0\gamma$ and $\rho^0\to\pi^0\gamma$ channels. These results show that the three-state limit does not provide an adequate description of the current radiative-decay data: constituent-quark mass differences alone are not sufficient, and an additional ingredient is required. In the conventional description, this ingredient is provided by the overlap factors; the extended framework tests whether the mixing with $X_0$ can provide it instead.

\subsection{Extended $SO(4)$ Fit}

We now perform the fit within the extended $SO(4)$ framework by allowing the additional mixing angles $\phi_{14}$, $\phi_{24}$ and $\phi_{34}$ to vary freely. For the ten-channel data set, the fit yields $\chi^2_{\rm min}/{\rm d.o.f.}=2.1/1$, compared with $26.6/4$ in the three-state limit, i.e.\ a reduction $\Delta\chi^2=24.5$. This improvement can be quantified with a likelihood-ratio test, since the three-state model is a special case of the extended one, and corresponds to a significance of $4.3\sigma$.\footnote{If the extra parameters were not needed, $\Delta\chi^2$ would follow a $\chi^2$ distribution with three degrees of freedom, giving a probability of $2\times10^{-5}$ for $\Delta\chi^2\geq 24.5$, i.e.\ $4.3\sigma$. Unlike $\chi^2_{\rm min}/{\rm d.o.f.}$, which measures the goodness of each fit separately, $\Delta\chi^2$ allows quantifying the significance of the improvement.} 
The extended mixing scheme therefore provides a significantly better description of the radiative-decay data. All significances quoted below are obtained in the same way. The resulting best-fit parameter values and their correlations are collected in Tables~\ref{tab:fit} and~\ref{tab:corr}.

\begin{table}[t]
\centering
\setlength{\tabcolsep}{2.3pt}
\renewcommand{\aboverulesep}{0.5pt}
\renewcommand{\belowrulesep}{0.5pt}
\captionsetup{font=small, skip=2pt}
\caption{Correlation matrix of the parameters in Table~\ref{tab:fit}.}
\label{tab:corr}
\begin{tabular}{lccccccccc}
\toprule
 & $g$ & $\epsilon_{12}$ & $\epsilon_{13}$ & $\phi_{14}$ & $\phi_{23}$ & $\phi_{24}$ & $\phi_{34}$ & $\phi_V$ & $\bar{m}/m_s$ \\
\midrule
$g$             & $1$ & $0.02$ & $-0.17$ & $0.61$ & $-0.09$ & $0.66$ & $0.04$ & $-0.32$ & $-0.90$ \\[1pt]
$\epsilon_{12}$ &      & $1$ & $-0.71$ & $0.39$ & $-0.53$ & $-0.53$ & $0.60$ & $-0.70$ & $0.13$ \\[1pt]
$\epsilon_{13}$ &      &      & $1$ & $-0.38$ & $0.78$ & $0.29$ & $-0.80$ & $0.96$ & $-0.01$ \\[1pt]
$\phi_{14}$     &      &      &      & $1$ & $-0.31$ & $0.18$ & $0.30$ & $-0.53$ & $-0.49$ \\[1pt]
$\phi_{23}$     &      &      &      &      & $1$ & $0.48$ & $-0.94$ & $0.83$ & $-0.10$ \\[1pt]
$\phi_{24}$     &      &      &      &      &      & $1$ & $-0.52$ & $0.25$ & $-0.71$ \\[1pt]
$\phi_{34}$     &      &      &      &      &      &      & $1$ & $-0.86$ & $0.09$ \\[1pt]
$\phi_V$        &      &      &      &      &      &      &      & $1$ & $0.14$ \\[1pt]
$\bar{m}/m_s$   &      &      &      &      &      &      &      &      & $1$ \\
\bottomrule
\end{tabular}
\end{table}

The uncertainties of the fit parameters are one-standard-deviation errors derived from the covariance matrix at the minimum of the $\chi^2$ function; for the three new mixing angles, the last column of Table~\ref{tab:fit} also gives the intervals obtained from the profile likelihood. Some parameters are strongly correlated, as can be understood from Eq.~(\ref{eq:couplings}).\footnote{The main correlations arise from combinations of parameters fixed by the data: the precise $\phi\to\pi^0\gamma$ width fixes $\sin\phi_V-\frac23(\bar m/m_s)\,\epsilon_{13}\cos\phi_V$; the strange amplitudes of the $\eta$ and $\eta^\prime$ depend on $\phi_{23}$ and $\phi_{34}$ essentially through $\phi_{23}+\beta$, with $\tan\beta=\sin\phi_{24}\tan\phi_{34}$ (see Sec.~\ref{sec:overlap}); and the $\phi\to\eta^{(\prime)}\gamma$ channels fix the product $g\,\bar m/m_s$.}
Our conclusions, however, rest on the flavour direction $\alpha$ and on $\chi^2$ differences between fits, which are obtained with all other parameters refitted and therefore account for these correlations.

Including the $K^*\to K\gamma$ decays in the fit gives $\chi^2_{\rm min}/{\rm d.o.f.}=17.7/3\simeq5.9$. The poorer fit quality originates in the kaon sector. The ratio of the two kaon couplings, $g_{K^{*+}K^+\gamma}/g_{K^{*0}K^0\gamma}=-(2-\bar m/m_s)/(1+\bar m/m_s)$, depends only on $\bar m/m_s$ and therefore fixes $\bar m/m_s=0.83\pm0.05$, about $4\sigma$ above the value preferred by the ten channels without kaons (see Table~\ref{tab:fit}). Since $\bar m/m_s$ enters all the couplings, the global fit accommodates the kaon data by raising $\bar m/m_s$ to essentially this value, and compensates the resulting enhancement of the strange contributions to the non-kaon couplings with a predominantly strange admixture, $\alpha\simeq78^\circ$, which suppresses the strange amplitudes. In the conventional description, this tension is instead absorbed by kaon-specific overlap factors~\cite{Escribano:2020jdy}, which have no counterpart in the extended scheme. The kaon decays thus probe $SU(3)_F$ breaking in the kaon sector rather than the neutral pseudoscalar mixing studied here, and we regard the ten-channel fit as the primary result of this analysis. Improved measurements of the $K^*\to K\gamma$ widths would clarify whether the tension is of experimental origin or signals additional $SU(3)_F$ breaking specific to the kaons.

\subsection{Structure of the Best-Fit Solution}
\label{sec:pattern}

The three additional mixing angles show a clear pattern (see Table~\ref{tab:fit}). The angle $\phi_{24}$, which describes the mixing of $X_0$ with the non-strange component, is clearly different from zero: fixing it to zero increases $\chi^2_{\rm min}$ by $15.0$, so that $\phi_{24}=0$ is disfavoured at the $3.9\sigma$ level (see Table~\ref{tab:nested}). In contrast, $\phi_{34}$ and $\phi_{14}$ are compatible with zero: fixing them to zero increases $\chi^2_{\rm min}$ by only $1.5$ and $1.8$, respectively $(1.2\sigma$ and $1.3\sigma)$. The data therefore indicate that the dominant effect of the extended mixing scheme arises from mixing between $X_0$ and the non-strange component $\eta_{\rm NS}$.

\begin{table}[t]
\centering
\setlength{\tabcolsep}{4pt}
\renewcommand{\aboverulesep}{0.5pt}
\renewcommand{\belowrulesep}{0.5pt}
\captionsetup{font=small, skip=2pt}
\caption{Minimum $\chi^2$ of the extended fit of Table~\ref{tab:fit} and of fits with restricted $X_0$ mixing, with $\Delta\chi^2$ relative to the former. In the third block, $X_0$ mixes only with the indicated component.}
\label{tab:nested}
\begin{tabular}{lcc}
\toprule
Fit & $\chi^2_{\rm min}$/d.o.f. & $\Delta\chi^2$ \\[0.5pt]
\midrule
Extended $SO(4)$ fit & $2.1/1$ & -- \\[1pt]
\midrule
$\phi_{14}=0$ & $3.9/2$ & $1.8$ \\[1pt]
$\phi_{34}=0$ & $3.6/2$ & $1.5$ \\[1pt]
$\phi_{24}=0$ & $17.1/2$ & $15.0$ \\[1pt]
\midrule
Non-strange mixing only $(\alpha=0)$ & $4.9/3$ & $2.7$ \\[1pt]
Singlet mixing only $(\alpha=35.3^\circ)$ & $8.3/3$ & $6.2$ \\[1pt]
Strange mixing only $(\alpha=90^\circ)$ & $26.6/3$ & $24.5$ \\[1pt]
\midrule
Three-state limit & $26.6/4$ & $24.5$ \\
\bottomrule
\end{tabular}
\end{table}

\begin{figure}[t]
\centering
\includegraphics[width=\columnwidth]{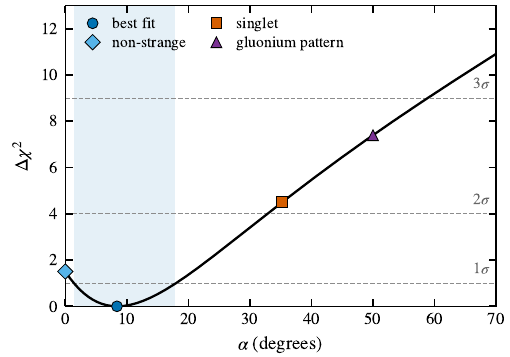}
\caption{Profile of $\Delta\chi^2$ as a function of the flavour direction $\alpha$ of the admixture, with all other parameters refitted. The shaded band is the $1\sigma$ interval. The markers indicate the purely non-strange and flavour-singlet directions, and the direction implied by the gluonium pattern of previous analyses, discussed in Sec.~\ref{sec:gluonium}.}
\label{fig:alpha}
\end{figure}

Since the individual angles depend on the parametrisation, this conclusion is best expressed in terms of the basis-independent angle $\alpha$ of Eq.~(\ref{eq:alpha}), which characterises the flavour direction of the admixture. The profile likelihood, shown in Fig.~\ref{fig:alpha}, gives $\alpha=(8.4^{+9.5}_{-7.0})^\circ$, close to the purely non-strange direction, $\alpha=0$, whereas the flavour-singlet direction, $\alpha=35.3^\circ$, is disfavoured at the $2.1\sigma$ level.

The same conclusion follows from fits in which the admixture is restricted to a single flavour direction, shown in the third block of Table~\ref{tab:nested}. Since these fits have the same number of parameters, their $\chi^2_{\rm min}$ values can be compared directly. A purely non-strange admixture describes the data well, with $\chi^2_{\rm min}/{\rm d.o.f.}=4.9/3$, whereas a flavour-singlet admixture gives $8.3/3$ and a purely strange one brings no improvement over the three-state limit. Moreover, the purely non-strange case differs from the extended fit by only $\Delta\chi^2=2.7$ with two fewer parameters ($1.1\sigma$), so that the data require a single additional parameter, the non-strange mixing angle $\phi_{24}$. As shown in Sec.~\ref{sec:overlap}, this case coincides exactly with the conventional overlap-factor model.

A more intuitive picture of the mixing structure is obtained from the squared elements of the rotation matrix, which quantify the fractional contribution of each basis state to the physical pseudoscalars. As shown in Table~\ref{tab:contributions}, the physical $\eta^\prime$ contains a $(12.4\pm2.7)\%$ admixture of $X_0$. Because the admixture is non-strange, the physical $\eta$ carries a comparable $X_0$ component, $(9.2\pm4.3)\%$, indicating that the effect of the extended mixing is not confined to the $\eta^\prime$ alone. The physical $\pi^0$ remains essentially pure $\pi_3$, while the fourth physical state $X$ is dominated by $X_0$ and contains an $\eta_{\rm NS}$ component of $(21.2\pm5.0)\%$ and an $\eta_{\rm S}$ component compatible with zero, $(0.5\pm0.8)\%$. This is the flavour pattern of a non-strange admixture. For a flavour-singlet admixture, the quark--antiquark part of $X$ would instead point along $\eta_0=(\sqrt2\,\eta_{\rm NS}+\eta_{\rm S})/\sqrt3$, with $\eta_{\rm NS}$ and $\eta_{\rm S}$ components in the ratio $2:1$.

Despite the introduction of $X_0$, the conventional mixing angles are only mildly affected. The $\eta$--$\eta^\prime$ mixing angle, $\phi_{23}=(39.9\pm1.3)^\circ$, is consistent with previous phenomenological determinations~\cite{Bramon:2000fr,Escribano:2007cd,Ambrosino:2009sc,Escribano:2020jdy}, with the chiral-Lagrangian prediction $\phi_{23}=41.4^\circ$ of Ref.~\cite{Mathieu:2010ss}, and with the most recent lattice determination, $\phi=(39.3\pm2.0)^\circ$~\cite{Ottnad:2025zxq}. Likewise, the vector mixing angle, $\phi_V=(3.7\pm0.5)^\circ$, remains fully consistent with the established picture of nearly ideal $\omega$--$\phi$ mixing. The isospin-breaking parameters are of a few percent, in line with Ref.~\cite{Escribano:2020jdy}, where their size relative to theoretical expectations is discussed in detail.

\begin{table}[t]
\centering
\setlength{\tabcolsep}{3.5pt}
\renewcommand{\aboverulesep}{0.5pt}
\renewcommand{\belowrulesep}{0.5pt}
\captionsetup{font=small, skip=2pt}
\caption{Composition (in \%) of the physical pseudoscalar states in the extended fit. Row sums differ from $100\%$ by at most $0.2\%$ owing to the first-order treatment of $\epsilon_{12}$ and $\epsilon_{13}$.}
\label{tab:contributions}
\begin{tabular}{lcccc}
\toprule
State & $\pi_3$ & $\eta_{\rm NS}$ & $\eta_{\rm S}$ & $X_0$ \\
\midrule
$\pi^0$       & $99.3 \pm 1.0$  & $0.01 \pm 0.04$ & $0.01 \pm 0.05$ & $0.9 \pm 1.0$  \\[1pt]
$\eta$        & $0.02 \pm 0.05$ & $46.4 \pm 4.2$  & $44.4 \pm 1.0$  & $9.2 \pm 4.3$  \\[1pt]
$\eta^\prime$ & $0.14 \pm 0.05$ & $32.5 \pm 1.9$  & $55.1 \pm 1.5$ & $12.4 \pm 2.7$ \\[1pt]
$X$           & $0.7 \pm 1.0$   & $21.2 \pm 5.0$  & $0.5 \pm 0.8$ & $77.6 \pm 4.9$ \\
\bottomrule
\end{tabular}
\end{table}

Table~\ref{tab:widths} compares the experimental and predicted radiative decay widths channel by channel. The improvement over the three-state limit is distributed across several channels rather than driven by a single observable: the tensions in the \mbox{$\omega\to\pi^0\gamma$}, $\eta^\prime\to\rho^0\gamma$ and $\rho^0\to\pi^0\gamma$ channels are removed,
and no channel deviates from its measured value by more than $1.3\sigma$. The largest remaining contribution to $\chi^2_{\rm min}$ comes from the $\rho^+\to\pi^+\gamma$ channel, which in practice controls the goodness of the extended fit. When this channel is excluded, the fit has as many parameters as observables, and the model predicts ${\rm BR}(\rho^+\to\pi^+\gamma)=(5.35\pm0.27)\times10^{-4}$. This value lies $1.5\sigma$ above the PDG average, $(4.5\pm0.5)\times10^{-4}$, which carries a scale factor $S=2.2$ and is based on measurements from the 1980s. A new measurement of this channel would therefore provide a direct test of the extended fit.

\begin{table*}[t]
\centering
\setlength{\tabcolsep}{6pt}
\renewcommand{\aboverulesep}{0.5pt}
\renewcommand{\belowrulesep}{0.5pt}
\captionsetup{font=small, skip=2pt}
\caption{Experimental radiative decay widths compared with the predictions of the three-state limit and the extended $SO(4)$ fit. The last two columns give the contribution of each channel to $\chi^2_{\rm min}$.}
\label{tab:widths}
\begin{tabular}{l r@{$\,\pm\,$}l r@{$\,\pm\,$}l r@{$\,\pm\,$}l cc}
\toprule
Channel
& \multicolumn{2}{c}{$\Gamma_{\rm exp}$ (keV)}
& \multicolumn{2}{c}{$\Gamma_{\rm 3\,states}$ (keV)}
& \multicolumn{2}{c}{$\Gamma_{\rm extended}$ (keV)}
& $\chi_i^2$ (3 states)
& $\chi_i^2$ (extended)
\\[1pt]
\midrule
$\rho^0\to\eta\gamma$        & $44$   & $3$    & $47$   & $1$    & $44$   & $3$    & 0.670 & 0.002 \\[1pt]
$\rho^0\to\pi^0\gamma$       & $69$   & $12$   & $89$   & $4$    & $74$   & $12$   & 2.37  & 0.147 \\[1pt]
$\rho^+\to\pi^+\gamma$       & $67$   & $7$    & $66$   & $1$    & $77$   & $3$    & 0.017 & 1.745 \\[1pt]
$\omega\to\eta\gamma$        & $3.9$  & $0.4$  & $4.1$  & $0.3$  & $3.9$  & $0.4$  & 0.170 & 0.001 \\[1pt]
$\omega\to\pi^0\gamma$       & $723$  & $24$   & $637$  & $15$   & $711$  & $23$   & 13.47 & 0.242 \\[1pt]
$\phi\to\eta\gamma$          & $55.5$ & $1.0$  & $54.8$ & $1.0$  & $55.5$ & $1.0$  & 0.468 & $<10^{-5}$ \\[1pt]
$\phi\to\eta^\prime\gamma$   & $0.26$ & $0.01$ & $0.28$ & $0.01$ & $0.26$ & $0.01$ & 1.48  & $<10^{-5}$ \\[1pt]
$\phi\to\pi^0\gamma$         & $5.7$  & $0.2$  & $5.7$  & $0.2$  & $5.7$  & $0.2$  & 0.025 & $<10^{-5}$ \\[1pt]
$\eta^\prime\to\rho^0\gamma$ & $55$   & $2$    & $61$   & $1$    & $55$   & $2$    & 7.69  & $<10^{-3}$ \\[1pt]
$\eta^\prime\to\omega\gamma$ & $4.7$  & $0.2$  & $4.6$  & $0.2$  & $4.7$  & $0.2$  & 0.271 & $<10^{-3}$ \\
\midrule
\multicolumn{7}{r}{Total $\chi^2_{\rm min}$/d.o.f.} & $26.6/4$ & $2.1/1$ \\[0.5pt]
\bottomrule
\end{tabular}
\end{table*}

\subsection{Correspondence with the Overlap-Factor Model}
\label{sec:overlap}

In the conventional description of Refs.~\cite{Bramon:2000fr,Escribano:2020jdy}, the second source of $SU(3)_F$ breaking is introduced through overlap factors that multiply the non-strange and strange contributions to the couplings. In the notation of Eq.~(\ref{eq:couplings}), the couplings of this model are obtained with the replacements $a_P^{\pi}\to\bar a_P^{\pi}$, $a_P^{\rm NS}\to z_{\rm NS}\,\bar a_P^{\rm NS}$ and $a_P^{\rm S}\to z_{\rm S}\,\bar a_P^{\rm S}$, where $\bar a_P^b\equiv a_P^b\big|_{\phi_{14}=\phi_{24}=\phi_{34}=0}$ are the amplitudes of the three-state limit.\footnote{In the isospin limit, for instance, where $\bar a_\eta^{\rm NS}=\cos\phi_{23}$, this gives $g_{\rho^0\eta\gamma}=g\,z_{\rm NS}\cos\phi_{23}$, the familiar result of Ref.~\cite{Bramon:2000fr}.} In the channels without kaons, $z_{\rm S}$ only appears in the product $z_{\rm S}\bar m/m_s$, so that $z_{\rm S}$ and $\bar m/m_s$ cannot be determined separately.

Within the extended scheme, the data are well described when $X_0$ mixes only with the non-strange component (see Sec.~\ref{sec:pattern}). This purely non-strange case is identical to the overlap-factor model: for $\phi_{14}=\phi_{34}=0$, the rotation matrix of Appendix~\ref{app:matrix} gives, to first order in $\epsilon_{12}$ and $\epsilon_{13}$,
\begin{equation}
a_P^{\rm NS}=\cos\phi_{24}\,\bar a_P^{\rm NS}\, ,\qquad
a_P^{\pi,{\rm S}}=\bar a_P^{\pi,{\rm S}}\, ,
\label{eq:NSonly}
\end{equation}
which reproduces the overlap-factor model with
\begin{equation}
z_{\rm NS}=\cos\phi_{24}\, ,\qquad z_{\rm S}=1\, .
\label{eq:zNS}
\end{equation}

Consequently, the fit of the purely non-strange case, with $\chi^2_{\rm min}/{\rm d.o.f.}=4.9/3$, is also the fit of the overlap-factor model to the current data. It gives $z_{\rm NS}=0.883\pm0.023$, $\phi_{23}=(41.4\pm0.5)^\circ$ and $z_{\rm S}\bar m/m_s=0.649\pm0.013$, in agreement with the values $z_{\rm NS}=0.89\pm0.03$, $\phi_{23}=(41.5\pm0.5)^\circ$ and $z_{\rm S}\bar m/m_s=0.65\pm0.01$ obtained in Fit 5 of Ref.~\cite{Escribano:2020jdy} with earlier data. The extended scheme gives this overlap factor a direct physical meaning: $z_{\rm NS}^2\simeq0.78$ is the fraction of $\eta_{\rm NS}$ that remains in the $\eta$--$\eta^\prime$ system, the remaining $22\%$ being carried by the fourth state $X$.

When $X_0$ is also allowed to mix with $\eta_{\rm S}$ and $\pi_3$, the correspondence is no longer exact, but it remains transparent. Neglecting small terms of order $\epsilon_{1j}\sin\phi_{14}$, the non-strange and strange amplitudes of the $\eta$ and $\eta^\prime$ (see Appendix~\ref{app:matrix}) can be written as
\[
\begin{aligned}
a_\eta^{\rm NS}&=z_{\rm NS}\cos\phi_{23}\, , & a_\eta^{\rm S}&=-z_{\rm S}\sin(\phi_{23}+\beta)\, ,\\
a_{\eta^\prime}^{\rm NS}&=z_{\rm NS}\sin\phi_{23}\, , & a_{\eta^\prime}^{\rm S}&=z_{\rm S}\cos(\phi_{23}+\beta)\, ,
\end{aligned}
\]
with
\begin{equation}
\begin{aligned}
z_{\rm NS}&=\cos\phi_{24}\, ,\\
z_{\rm S}&=\sqrt{1-\cos^2\phi_{24}\sin^2\phi_{34}}\, ,\\[2pt]
\tan \beta&=\sin\phi_{24}\tan\phi_{34}\, .
\end{aligned}
\label{eq:zgeneral}
\end{equation}
The non-strange amplitudes are rescaled exactly as in Eq.~(\ref{eq:NSonly}), while the strange ones are rescaled by $z_{\rm S}$ and shifted by the angle $\beta$, which is the only difference with respect to the overlap-factor model.\footnote{The $\pi^0$ amplitudes have a similar structure. Writing $(\epsilon_{13},\epsilon_{12})=\epsilon\,(\cos\chi,\sin\chi)$, one finds $a_{\pi^0}^{\rm NS}=\epsilon\,z_{\rm NS}\sin\chi-\sin\phi_{14}\sin\phi_{24}$ and $a_{\pi^0}^{\rm S}=\epsilon\,z_{\rm S}\cos(\chi+\beta)-\sin\phi_{14}\cos\phi_{24}\sin\phi_{34}$.} At the best-fit point of the extended fit, \mbox{$z_{\rm NS}=0.887\pm0.028$}, $z_{\rm S}=0.998\pm0.004$ and $\beta=(2.1\pm1.7)^\circ$, so that it reproduces the overlap-factor phenomenology with $z_{\rm S}\simeq1$ and a small shift $\beta$.


\section{Physical Interpretation}
\label{sec:interpretation}

The fit results of Sec.~\ref{sec:fit} show that extending the pseudoscalar mixing sector by a single additional state substantially improves the description of the radiative-decay data relative to the three-state limit, and that the admixture favoured by the data is predominantly non-strange. Although the fit cannot identify the microscopic nature of $X_0$, this flavour structure provides clear phenomenological guidance for its interpretation. In the following, we first confront it with the gluonium hypothesis that has motivated much of the previous work, and then discuss the interpretation in terms of radially excited states and, briefly, in terms of a fundamental pseudoscalar particle.

\subsection{Gluonium in the $\eta^\prime$ and the Pseudoscalar Glueball}
\label{sec:gluonium}

Mixing between a pseudoscalar glueball and the quark--antiquark states is a generic expectation of the axial $U(1)_A$ anomaly, which appears in the divergence of the flavour-singlet axial current~\cite{Adler:1969gk,Bell:1969ts,Fujikawa:1979ay},
\[
\partial_\mu J_5^\mu
=
2i\!\sum_q m_q\,\bar q\gamma_5 q
+
\frac{N_f\alpha_s}{4\pi}
G_{\mu\nu}^a\tilde G^{a\,\mu\nu}\, ,
\]
where $J_5^\mu=\sum_q\bar q\gamma^\mu\gamma_5 q$ and $\tilde G^{a\,\mu\nu}=\tfrac12\epsilon^{\mu\nu\rho\sigma}G^a_{\rho\sigma}$. The second, purely gluonic term on the right-hand side is the operator through which such a state couples to the quark--antiquark sector. Since this operator is flavour blind, the resulting admixture is expected along the flavour-singlet direction $\eta_0$.\footnote{$SU(3)_F$ breaking may modify this expectation. Weighting the anomaly couplings by the decay constants, as in the Feldmann--Kroll--Stech scheme~\cite{Feldmann:1998vh}, shifts the expected direction to $\tan\alpha=f_q/(\sqrt2 f_s)$, i.e.\ $\alpha\simeq30^\circ$ for $f_q/f_s\simeq0.81$, which remains disfavoured by the data at the $1.8\sigma$ level.} As shown in Sec.~\ref{sec:pattern}, this direction is disfavoured by the data at the $2.1\sigma$ level.

The gluonium pattern assumed in previous analyses~\cite{Escribano:2007cd,KLOE:2006guu,Ambrosino:2009sc}, with a glue-free $\eta$ and the whole gluonic admixture in the $\eta^\prime$, fares no better. Since the $\eta$ must then remain free of $X_0$, the admixture has to lie along the quark--antiquark part of the $\eta^\prime$, at $\alpha\simeq50^\circ$,\footnote{Precisely, $\alpha=90^\circ-\phi_P$, where $\phi_P$ is the $\eta$--$\eta^\prime$ mixing angle defined, as in the KLOE analyses, by the quark--antiquark content of the $\eta$; it differs from $\phi_{23}$ when $X_0$ mixes with the $\eta$ and $\eta^\prime$. Imposing a glue-free $\eta$, we obtain $\phi_P=(40.0\pm0.5)^\circ$, in agreement with the KLOE values $39.7^\circ$--$40.4^\circ$~\cite{KLOE:2006guu,Ambrosino:2009sc}.} which lies far from the preferred direction, $\alpha=(8.4^{+9.5}_{-7.0})^\circ$, and is disfavoured at the $2.7\sigma$ level (see Fig.~\ref{fig:alpha}). Imposing the pattern nevertheless reproduces the KLOE result, with an $X_0$ content of the $\eta^\prime$ of $(13.8\pm3.2)\%$, compared with $Z_{\eta^\prime}^2=0.12$--$0.14$~\cite{KLOE:2006guu,Ambrosino:2009sc}. The gluonium fraction found in those analyses therefore measures the size of the suppression required by the data, while its identification with gluonium is an assumption rather than a result.

Moreover, the size of the admixture is hard to reconcile with a heavy glueball. In a two-state picture, an admixture angle $\theta$ between the $\eta^\prime$ and a state of mass $m_X$ requires an off-diagonal element of the squared-mass matrix, $\delta M^2=\tfrac12(m_X^2-m_{\eta^\prime}^2)\sin2\theta$. For the $X_0$ content of the $\eta^\prime$ found in the extended fit, $\sin^2\theta=12.4\%$, i.e.\ $\theta\simeq21^\circ$, a glueball in the lattice mass range $2.3$--$3.0$~GeV~\cite{Chen:2005mg,Gregory:2012hu,Athenodorou:2020ani} would require $\delta M^2\simeq1.4$--$2.7~{\rm GeV}^2$, four to eight times the value $\delta M^2\simeq0.3~{\rm GeV}^2$ found in the lattice study of glueball--singlet mixing of Ref.~\cite{Jiang:2022ffl}.\footnote{The lattice calculation is performed in $N_f=2$ QCD, where the $\eta$ is the $SU(2)$ flavour singlet and thus plays the role of the real-world $\eta^\prime$. It uses $m_\pi\approx350$~MeV, a stable glueball and no continuum or chiral extrapolation. Its off-diagonal linear mass-matrix element, $|x|=107(15)(2)$~MeV, corresponds to $\delta M^2\simeq|x|(m_G+m_\eta)\simeq0.3~{\rm GeV}^2$ in our convention. A chiral effective Lagrangian gives larger values, $\delta M^2\simeq1.7$--$2.1~{\rm GeV}^2$~\cite{Mathieu:2009sg}, but we rely on the first-principles lattice result, which we use as an order-of-magnitude estimate.} Lighter glueball candidates discussed in the literature, such as the $\eta(1405)$ or the $\eta(1760)$~\cite{ParticleDataGroup:2026aaa}, would evade this argument, but not the flavour argument above, which does not depend on the glueball mass.

The apparent conflict between the gluonium fractions of order $10\%$ extracted from radiative decays and the small glueball--singlet mixing found on the lattice is thus removed: the data require a suppression of the quark--antiquark couplings of this size, but it need not be gluonic. The recent identification of the $\eta(2370)$ as a glueball-dominated state by the BESIII Collaboration~\cite{BESIII:2026mvn} fits well into this picture. This $0^{-+}$ isoscalar, observed in radiative $J/\psi$ decays~\cite{BESIII:2023wfi,Huang:2025pyv}, has a mass of $2317\pm9$~MeV~\cite{ParticleDataGroup:2026aaa}, close to the lattice predictions for the lightest pseudoscalar glueball~\cite{Chen:2005mg,Gregory:2012hu,Athenodorou:2020ani}.\footnote{Being a flavour singlet does not by itself single out a glueball interpretation; see Ref.~\cite{Zhang:2026poz} for a discussion of competing assignments. The $\eta(2370)$ also lies in a region where broad and overlapping radially excited $0^{-+}$ quark--antiquark states are expected~\cite{ParticleDataGroup:2026aaa}.} Given the small mixing found on the lattice, such a glueball is unlikely to leave a visible imprint on radiative decays.

\subsection{Radially Excited Non-Strange States}

The most natural interpretation of the non-strange admixture favoured by the data is that $X_0$ represents radially excited non-strange quark--antiquark configurations, a possibility already pointed out in Ref.~\cite{Thomas:2007uy}. In the quark model, the radial excitations of the isoscalar pseudoscalars are expected to be close to ideally mixed, i.e.\ almost purely non-strange and strange states~\cite{Godfrey:1985xj}, as suggested by the near mass degeneracy of the $\eta(1295)$ and the $\pi(1300)$. The $\eta(1295)$ and the $\eta(1475)$ are then the non-strange and strange members of the first radially excited pseudoscalar nonet~\cite{ParticleDataGroup:2026aaa,Cheng:2021nal}.\footnote{The existence of the $\eta(1295)$, very close in mass to the $f_1(1285)$, has been questioned~\cite{Klempt:2007cp}, and whether the $\eta(1440)$ is a single state or two distinct ones, the $\eta(1405)$ and the $\eta(1475)$, is still open~\cite{ParticleDataGroup:2026aaa}. Our argument relies only on the existence of radially excited non-strange configurations, not on their identification with a specific resonance. Conversely, the fact that radiative decays require a predominantly non-strange admixture provides independent, albeit indirect, support for the existence of a non-strange radial excitation such as the $\eta(1295)$.} Mixing with such a state would affect only the non-strange components of the $\eta$ and $\eta^\prime$, leaving the strange ones essentially untouched, exactly as the data require. The size of the admixture is also easier to accommodate in this picture: since the $\eta(1295)$ lies much closer to the $\eta^\prime$, the two-state estimate of Sec.~\ref{sec:gluonium} requires only $\delta M^2\simeq0.25~{\rm GeV}^2$, about an order of magnitude less than for a state in the $2.3$--$3.0$~GeV range.

This interpretation also provides a physical basis for the central assumption of our framework, namely that $X_0$ does not couple at the $V\!P\gamma$ vertex and contributes to radiative decays only through its admixture into the $\eta$ and $\eta^\prime$. A \mbox{$V\to P\gamma$} transition is a magnetic-dipole transition which, for small photon momenta, flips the spin of a quark but leaves its spatial wavefunction unchanged, so that its amplitude is proportional to the overlap of the radial wavefunctions of the two mesons. This overlap is close to unity for ground states, but vanishes for a ground-state vector and a radially excited pseudoscalar, whose radial wavefunctions are orthogonal. Such transitions arise only through small recoil and relativistic corrections and are strongly suppressed~\cite{Thomas:2007uy}, as observed for the analogous $2S\to1S$ transition $\psi(2S)\to\eta_c\gamma$ in charmonium~\cite{ParticleDataGroup:2026aaa}.

In this picture, $z_{\rm NS}$ has a simple interpretation: part of the non-strange content of the $\eta$ and $\eta^\prime$ is carried by the radially excited state, which does not overlap with the $1S$ vector mesons, so that the non-strange contributions to their couplings are reduced by $z_{\rm NS}=\cos\phi_{24}$. This is in line with the original motivation of the overlap factors, which were introduced to account for the different spatial extensions of the meson wavefunctions~\cite{Bramon:2000fr}. A dynamical treatment of the mixing between ground and radially excited pseudoscalars would allow $z_{\rm NS}$ to be predicted and compared with the value extracted here.

\subsection{Interpretation as a Fundamental Degree of Freedom}

A more speculative interpretation is that $X_0$ is a new light fundamental pseudoscalar, such as an axion or an axion-like particle (ALP)~\cite{Peccei:1977hh,Peccei:1977ur,Weinberg:1977ma,Wilczek:1977pj,Kim:1986ax,Cheng:1987gp,Kim:2008hd,Marsh:2015xka}. In many ALP models, the mixing with neutral pseudoscalar mesons scales inversely with the symmetry-breaking scale $f_a$~\cite{Alves:2024dpa}, so that an admixture of the size found here would require $f_a\sim f_\pi$, a value excluded by laboratory, flavour, astrophysical and cosmological observations~\cite{Depta:2020wmr,Bauer:2021mvw,Alves:2017avw,Cornella:2019uxs,Chang:2018rso,Raffelt:2006cw,Bauer:2017ris}. ALPs compatible with these bounds would mix with the $\eta$ and $\eta^\prime$ far below the sensitivity of radiative decays, which probe mixing angles of a few degrees. Moreover, an ALP coupled to gluons mixes through the anomaly and therefore, like a glueball, predominantly along the flavour-singlet direction disfavoured by the data (see Sec.~\ref{sec:gluonium}), while a non-strange admixture would require \textit{ad hoc} couplings to the light quarks. The present analysis therefore does not favour this interpretation.


\section{Conclusions}
\label{sec:conclusions}

This work explores whether radiative $V\to P\gamma$ and $P\to V\gamma$ decays can be described without phenomenological wavefunction-overlap factors by extending the pseudoscalar mixing sector. To this end, we enlarge the conventional three-state pseudoscalar basis with one additional state, $X_0$, which does not couple at the $V\!P\gamma$ vertex, and parametrise the mixing by an $SO(4)$ rotation in the quark-flavour basis, with the overlap factors fixed to unity throughout.

For the ten channels that do not involve kaons, the extended mixing improves the description with respect to the three-state limit with a significance of $4.3\sigma$, while preserving the established $\eta$--$\eta^\prime$ and vector mixing angles. The best-fit solution shows a clear pattern: $X_0$ mixes predominantly with the non-strange component $\eta_{\rm NS}$, with $\phi_{24}=(27.5\pm3.5)^\circ$, whereas its mixing with the strange component and with the pion is compatible with zero. Accordingly, the basis-independent direction of the admixture, $\alpha=(8.4^{+9.5}_{-7.0})^\circ$, is close to the purely non-strange one.

This pattern establishes a direct correspondence with the conventional description. When $X_0$ mixes only with the non-strange component, the extended scheme is identical, channel by channel, to the overlap-factor model of Ref.~\cite{Escribano:2020jdy}, with $z_{\rm NS}=\cos\phi_{24}=0.883\pm0.023$ for current data. The non-strange overlap factor thus acquires a precise physical meaning: $z_{\rm NS}^2$ is the fraction of the non-strange component that remains in the $\eta$--$\eta^\prime$ system.

These results also bear on the long-standing question of gluonium in the $\eta^\prime$. The data disfavour both a flavour-singlet admixture, as expected from anomaly-driven mixing with a pseudoscalar glueball, and the gluonium pattern assumed in previous analyses, at the $2.1\sigma$ and $2.7\sigma$ levels, respectively. Moreover, attributing the admixture to a glueball in the lattice mass range would require an off-diagonal mass-matrix element several times larger than the one found in lattice QCD~\cite{Jiang:2022ffl}. Radiative decays therefore provide no compelling evidence for gluonium in the $\eta^\prime$, a conclusion consistent with a weakly mixing glueball such as the $\eta(2370)$.

The non-strange pattern is instead naturally accounted for by radially excited non-strange configurations, such as the $\eta(1295)$. Such states require a much smaller off-diagonal mass-matrix element than a heavy glueball, and they are expected to contribute to radiative decays only through their mixing with the $\eta$ and $\eta^\prime$, since magnetic-dipole transitions between ground-state vectors and radially excited pseudoscalars are strongly suppressed. In this picture, the departure of $z_{\rm NS}$ from unity reflects the radially excited component of the $\eta$ and $\eta^\prime$, which does not couple directly to the ground-state vector mesons, in line with the original motivation of the overlap factors.

Two measurements would test the proposed framework: a new determination of $\rho^+\to\pi^+\gamma$, which controls the residual $\chi^2$ of the extended fit and for which the model predicts ${\rm BR}(\rho^+\to\pi^+\gamma)=(5.35\pm0.27)\times10^{-4}$, and improved $K^*\to K\gamma$ widths, which would clarify the origin of the tension in the kaon sector. On the theoretical side, a dynamical calculation of the mixing between ground-state and radially excited isoscalar pseudoscalars, in quark models or lattice QCD, would test the radial-excitation interpretation by predicting $z_{\rm NS}=\cos\phi_{24}$, to be compared with the value extracted here.


\section*{Acknowledgements}

The work of R.~E.~and E.~R.~is supported by the Ministerio de Ciencia, Innovaci{\'o}n y Universidades under Grant No.~PID2023-146142NB-I00.
The work of R.~E.~is also supported by the Spanish Ministry of Science and Innovation (MICINN) through the Spanish State Research Agency, under Severo Ochoa Centres of Excellence Programme 2025-2029 (CEX2024-001442-S). IFAE is partially funded by the CERCA program of the Generalitat de Catalunya.
The work of E.~R.~is additionally funded by the Universidad Cardenal Herrera-CEU (INDI25/17 and GIR25/15).


\appendix
\section{$SO(4)$ Rotation Matrix}
\label{app:matrix}

The rows of the rotation matrix $R$ of Eq.~(\ref{eq:Rdef}) correspond to the physical states $(\pi^0,\eta,\eta^\prime,X)$ and its columns to the basis states $(\pi_3,\eta_{\rm NS},\eta_{\rm S},X_0)$, as in Eq.~(\ref{eq:states}), so that the elements of the first three rows are the amplitudes of Eq.~(\ref{eq:components}), $R_{Pb}=a_P^b$. In the shorthand notation of Sec.~\ref{sec:model}, and to first order in $\epsilon_{12}$ and $\epsilon_{13}$ $(\sin\phi_{1j}\simeq\epsilon_{1j}$, $\cos\phi_{1j}\simeq1$, $j=2,3)$, its elements read
\begin{widetext}
\small
\begin{align*}
R_{11} &= c_{14}\, , &
R_{31} &= -c_{14}(\epsilon_{13}c_{23}+\epsilon_{12}s_{23})\, , \\
R_{12} &= \epsilon_{12}c_{24}-s_{14}s_{24}\, , &
R_{32} &= c_{24}s_{23}+s_{14}(\epsilon_{13}c_{23}+\epsilon_{12}s_{23})s_{24}\, , \\
R_{13} &= \epsilon_{13}c_{34}-(c_{24}s_{14}+\epsilon_{12}s_{24})s_{34}\, , &
R_{33} &= s_{23}(\epsilon_{12}c_{24}s_{14}-s_{24})s_{34}+c_{23}(c_{34}+\epsilon_{13}c_{24}s_{14}s_{34})\, , \\
R_{14} &= c_{24}c_{34}s_{14}+\epsilon_{12}c_{34}s_{24}+\epsilon_{13}s_{34}\, , &
R_{34} &= c_{34}s_{23}(-\epsilon_{12}c_{24}s_{14}+s_{24})+c_{23}(-\epsilon_{13}c_{24}c_{34}s_{14}+s_{34})\, , \\
R_{21} &= c_{14}(-\epsilon_{12}c_{23}+\epsilon_{13}s_{23})\, , &
R_{41} &= -s_{14}\, , \\
R_{22} &= -\epsilon_{13}s_{14}s_{23}s_{24}+c_{23}(c_{24}+\epsilon_{12}s_{14}s_{24})\, , &
R_{42} &= -c_{14}s_{24}\, , \\
R_{23} &= -c_{34}s_{23}-\big(\epsilon_{13}c_{24}s_{14}s_{23}+c_{23}(-\epsilon_{12}c_{24}s_{14}+s_{24})\big)s_{34}\, , &
R_{43} &= -c_{14}c_{24}s_{34}\, , \\
R_{24} &= c_{23}c_{34}(-\epsilon_{12}c_{24}s_{14}+s_{24})+s_{23}(\epsilon_{13}c_{24}c_{34}s_{14}-s_{34})\, , &
R_{44} &= c_{14}c_{24}c_{34}\, .
\end{align*}
\normalsize
\end{widetext}


\section{Explicit Radiative Decay Couplings}
\label{app:couplings}

Inserting the amplitudes of the physical pseudoscalars from Appendix~\ref{app:matrix} into Eq.~(\ref{eq:couplings}), and defining the combinations
\[
E_\eta\equiv\epsilon_{12}c_{23}-\epsilon_{13}s_{23}\, ,\qquad
E_{\eta^\prime}\equiv\epsilon_{13}c_{23}+\epsilon_{12}s_{23}\, ,
\]
which govern the isospin-breaking $\pi_3$ components of the $\eta$ and $\eta^\prime$, the couplings of the neutral channels read
\begin{widetext}
\small
\begin{align*}
g_{\rho^0\pi^0\gamma}&=g\left[\tfrac13\,c_{14}+\epsilon_{12}c_{24}-s_{14}s_{24}\right] ,\\
g_{\rho^0\eta\gamma}&=g\left[-\tfrac13\,c_{14}E_\eta+c_{23}c_{24}+s_{14}s_{24}E_\eta\right] ,\\
g_{\rho^0\eta^\prime\gamma}&=g\left[-\tfrac13\,c_{14}E_{\eta^\prime}+s_{23}c_{24}+s_{14}s_{24}E_{\eta^\prime}\right] ,\\[2mm]
g_{\omega\pi^0\gamma}&=g\left\{c_V\left[c_{14}+\tfrac13\left(\epsilon_{12}c_{24}-s_{14}s_{24}\right)\right]+\tfrac23\tfrac{\bar m}{m_s}\,s_V\left[\epsilon_{13}c_{34}-\left(c_{24}s_{14}+\epsilon_{12}s_{24}\right)s_{34}\right]\right\} ,\\
g_{\omega\eta\gamma}&=g\left\{c_V\left[-c_{14}E_\eta+\tfrac13\left(c_{23}c_{24}+s_{14}s_{24}E_\eta\right)\right]-\tfrac23\tfrac{\bar m}{m_s}\,s_V\left[s_{23}c_{34}+\left(c_{23}s_{24}-c_{24}s_{14}E_\eta\right)s_{34}\right]\right\} ,\\
g_{\omega\eta^\prime\gamma}&=g\left\{c_V\left[-c_{14}E_{\eta^\prime}+\tfrac13\left(s_{23}c_{24}+s_{14}s_{24}E_{\eta^\prime}\right)\right]+\tfrac23\tfrac{\bar m}{m_s}\,s_V\left[c_{23}c_{34}-\left(s_{23}s_{24}-c_{24}s_{14}E_{\eta^\prime}\right)s_{34}\right]\right\} ,\\[2mm]
g_{\phi\pi^0\gamma}&=g\left\{s_V\left[c_{14}+\tfrac13\left(\epsilon_{12}c_{24}-s_{14}s_{24}\right)\right]-\tfrac23\tfrac{\bar m}{m_s}\,c_V\left[\epsilon_{13}c_{34}-\left(c_{24}s_{14}+\epsilon_{12}s_{24}\right)s_{34}\right]\right\} ,\\
g_{\phi\eta\gamma}&=g\left\{s_V\left[-c_{14}E_\eta+\tfrac13\left(c_{23}c_{24}+s_{14}s_{24}E_\eta\right)\right]+\tfrac23\tfrac{\bar m}{m_s}\,c_V\left[s_{23}c_{34}+\left(c_{23}s_{24}-c_{24}s_{14}E_\eta\right)s_{34}\right]\right\} ,\\
g_{\phi\eta^\prime\gamma}&=g\left\{s_V\left[-c_{14}E_{\eta^\prime}+\tfrac13\left(s_{23}c_{24}+s_{14}s_{24}E_{\eta^\prime}\right)\right]-\tfrac23\tfrac{\bar m}{m_s}\,c_V\left[c_{23}c_{34}-\left(s_{23}s_{24}-c_{24}s_{14}E_{\eta^\prime}\right)s_{34}\right]\right\} .
\end{align*}
\normalsize
\end{widetext}
The charged and strange couplings are those of Eq.~(\ref{eq:charged}). In the $\omega$ and $\phi$ couplings, the first square bracket is the combination $a_P^{\pi}+\tfrac13 a_P^{\rm NS}$ of Eq.~(\ref{eq:couplings}), and the second is the strange amplitude $a_P^{\rm S}$, with the opposite sign for the $\eta$. In the three-state limit, all these couplings reduce to those of Ref.~\cite{Escribano:2020jdy} with the overlap factors set to unity.

\bibliography{refs}

\begin{thebibliography}{64}%
\makeatletter
\providecommand \@ifxundefined [1]{%
 \@ifx{#1\undefined}
}%
\providecommand \@ifnum [1]{%
 \ifnum #1\expandafter \@firstoftwo
 \else \expandafter \@secondoftwo
 \fi
}%
\providecommand \@ifx [1]{%
 \ifx #1\expandafter \@firstoftwo
 \else \expandafter \@secondoftwo
 \fi
}%
\providecommand \natexlab [1]{#1}%
\providecommand \enquote  [1]{``#1''}%
\providecommand \bibnamefont  [1]{#1}%
\providecommand \bibfnamefont [1]{#1}%
\providecommand \citenamefont [1]{#1}%
\providecommand \href@noop [0]{\@secondoftwo}%
\providecommand \href [0]{\begingroup \@sanitize@url \@href}%
\providecommand \@href[1]{\@@startlink{#1}\@@href}%
\providecommand \@@href[1]{\endgroup#1\@@endlink}%
\providecommand \@sanitize@url [0]{\catcode `\\12\catcode `\$12\catcode
  `\&12\catcode `\#12\catcode `\^12\catcode `\_12\catcode `\%12\relax}%
\providecommand \@@startlink[1]{}%
\providecommand \@@endlink[0]{}%
\providecommand \url  [0]{\begingroup\@sanitize@url \@url }%
\providecommand \@url [1]{\endgroup\@href {#1}{\urlprefix }}%
\providecommand \urlprefix  [0]{URL }%
\providecommand \Eprint [0]{\href }%
\providecommand \doibase [0]{http://dx.doi.org/}%
\providecommand \selectlanguage [0]{\@gobble}%
\providecommand \bibinfo  [0]{\@secondoftwo}%
\providecommand \bibfield  [0]{\@secondoftwo}%
\providecommand \translation [1]{[#1]}%
\providecommand \BibitemOpen [0]{}%
\providecommand \bibitemStop [0]{}%
\providecommand \bibitemNoStop [0]{.\EOS\space}%
\providecommand \EOS [0]{\spacefactor3000\relax}%
\providecommand \BibitemShut  [1]{\csname bibitem#1\endcsname}%
\let\auto@bib@innerbib\@empty
\bibitem [{\citenamefont {Adler}(1969)}]{Adler:1969gk}%
  \BibitemOpen
  \bibfield  {author} {\bibinfo {author} {\bibfnamefont {S.~L.}\ \bibnamefont
  {Adler}},\ }\href {\doibase 10.1103/PhysRev.177.2426} {\bibfield  {journal}
  {\bibinfo  {journal} {Phys. Rev.}\ }\textbf {\bibinfo {volume} {177}},\
  \bibinfo {pages} {2426} (\bibinfo {year} {1969})}\BibitemShut {NoStop}%
\bibitem [{\citenamefont {Bell}\ and\ \citenamefont
  {Jackiw}(1969)}]{Bell:1969ts}%
  \BibitemOpen
  \bibfield  {author} {\bibinfo {author} {\bibfnamefont {J.~S.}\ \bibnamefont
  {Bell}}\ and\ \bibinfo {author} {\bibfnamefont {R.}~\bibnamefont {Jackiw}},\
  }\href {\doibase 10.1007/BF02823296} {\bibfield  {journal} {\bibinfo
  {journal} {Nuovo Cim. A}\ }\textbf {\bibinfo {volume} {60}},\ \bibinfo
  {pages} {47} (\bibinfo {year} {1969})}\BibitemShut {NoStop}%
\bibitem [{\citenamefont {Fujikawa}(1979)}]{Fujikawa:1979ay}%
  \BibitemOpen
  \bibfield  {author} {\bibinfo {author} {\bibfnamefont {K.}~\bibnamefont
  {Fujikawa}},\ }\href {\doibase 10.1103/PhysRevLett.42.1195} {\bibfield
  {journal} {\bibinfo  {journal} {Phys. Rev. Lett.}\ }\textbf {\bibinfo
  {volume} {42}},\ \bibinfo {pages} {1195} (\bibinfo {year}
  {1979})}\BibitemShut {NoStop}%
\bibitem [{\citenamefont {Witten}(1979)}]{Witten:1979vv}%
  \BibitemOpen
  \bibfield  {author} {\bibinfo {author} {\bibfnamefont {E.}~\bibnamefont
  {Witten}},\ }\href {\doibase 10.1016/0550-3213(79)90031-2} {\bibfield
  {journal} {\bibinfo  {journal} {Nucl. Phys. B}\ }\textbf {\bibinfo {volume}
  {156}},\ \bibinfo {pages} {269} (\bibinfo {year} {1979})}\BibitemShut
  {NoStop}%
\bibitem [{\citenamefont {Veneziano}(1979)}]{Veneziano:1979ec}%
  \BibitemOpen
  \bibfield  {author} {\bibinfo {author} {\bibfnamefont {G.}~\bibnamefont
  {Veneziano}},\ }\href {\doibase 10.1016/0550-3213(79)90332-8} {\bibfield
  {journal} {\bibinfo  {journal} {Nucl. Phys. B}\ }\textbf {\bibinfo {volume}
  {159}},\ \bibinfo {pages} {213} (\bibinfo {year} {1979})}\BibitemShut
  {NoStop}%
\bibitem [{\citenamefont {Cichy}\ \emph {et~al.}(2015)\citenamefont {Cichy},
  \citenamefont {Garcia-Ramos}, \citenamefont {Jansen}, \citenamefont
  {Ottnad},\ and\ \citenamefont {Urbach}}]{Cichy:2015jra}%
  \BibitemOpen
  \bibfield  {author} {\bibinfo {author} {\bibfnamefont {K.}~\bibnamefont
  {Cichy}}, \bibinfo {author} {\bibfnamefont {E.}~\bibnamefont {Garcia-Ramos}},
  \bibinfo {author} {\bibfnamefont {K.}~\bibnamefont {Jansen}}, \bibinfo
  {author} {\bibfnamefont {K.}~\bibnamefont {Ottnad}}, \ and\ \bibinfo {author}
  {\bibfnamefont {C.}~\bibnamefont {Urbach}} (\bibinfo {collaboration} {ETM}),\
  }\href {\doibase 10.1007/JHEP09(2015)020} {\bibfield  {journal} {\bibinfo
  {journal} {JHEP}\ }\textbf {\bibinfo {volume} {09}},\ \bibinfo {pages} {020}
  (\bibinfo {year} {2015})},\ \Eprint {http://arxiv.org/abs/1504.07954}
  {arXiv:1504.07954 [hep-lat]} \BibitemShut {NoStop}%
\bibitem [{\citenamefont {Feldmann}(2000)}]{Feldmann:1999uf}%
  \BibitemOpen
  \bibfield  {author} {\bibinfo {author} {\bibfnamefont {T.}~\bibnamefont
  {Feldmann}},\ }\href {\doibase 10.1142/S0217751X00000082} {\bibfield
  {journal} {\bibinfo  {journal} {Int. J. Mod. Phys. A}\ }\textbf {\bibinfo
  {volume} {15}},\ \bibinfo {pages} {159} (\bibinfo {year} {2000})},\ \Eprint
  {http://arxiv.org/abs/hep-ph/9907491} {arXiv:hep-ph/9907491} \BibitemShut
  {NoStop}%
\bibitem [{\citenamefont {Bass}\ and\ \citenamefont
  {Moskal}(2019)}]{Bass:2018xmz}%
  \BibitemOpen
  \bibfield  {author} {\bibinfo {author} {\bibfnamefont {S.~D.}\ \bibnamefont
  {Bass}}\ and\ \bibinfo {author} {\bibfnamefont {P.}~\bibnamefont {Moskal}},\
  }\href {\doibase 10.1103/RevModPhys.91.015003} {\bibfield  {journal}
  {\bibinfo  {journal} {Rev. Mod. Phys.}\ }\textbf {\bibinfo {volume} {91}},\
  \bibinfo {pages} {015003} (\bibinfo {year} {2019})},\ \Eprint
  {http://arxiv.org/abs/1810.12290} {arXiv:1810.12290 [hep-ph]} \BibitemShut
  {NoStop}%
\bibitem [{\citenamefont {Gan}\ \emph {et~al.}(2022)\citenamefont {Gan},
  \citenamefont {Kubis}, \citenamefont {Passemar},\ and\ \citenamefont
  {Tulin}}]{Gan:2020aco}%
  \BibitemOpen
  \bibfield  {author} {\bibinfo {author} {\bibfnamefont {L.}~\bibnamefont
  {Gan}}, \bibinfo {author} {\bibfnamefont {B.}~\bibnamefont {Kubis}}, \bibinfo
  {author} {\bibfnamefont {E.}~\bibnamefont {Passemar}}, \ and\ \bibinfo
  {author} {\bibfnamefont {S.}~\bibnamefont {Tulin}},\ }\href {\doibase
  10.1016/j.physrep.2021.11.001} {\bibfield  {journal} {\bibinfo  {journal}
  {Phys. Rept.}\ }\textbf {\bibinfo {volume} {945}},\ \bibinfo {pages} {1}
  (\bibinfo {year} {2022})},\ \Eprint {http://arxiv.org/abs/2007.00664}
  {arXiv:2007.00664 [hep-ph]} \BibitemShut {NoStop}%
\bibitem [{\citenamefont {Ball}\ \emph {et~al.}(1996)\citenamefont {Ball},
  \citenamefont {Frere},\ and\ \citenamefont {Tytgat}}]{Ball:1995zv}%
  \BibitemOpen
  \bibfield  {author} {\bibinfo {author} {\bibfnamefont {P.}~\bibnamefont
  {Ball}}, \bibinfo {author} {\bibfnamefont {J.~M.}\ \bibnamefont {Frere}}, \
  and\ \bibinfo {author} {\bibfnamefont {M.}~\bibnamefont {Tytgat}},\ }\href
  {\doibase 10.1016/0370-2693(95)01287-7} {\bibfield  {journal} {\bibinfo
  {journal} {Phys. Lett. B}\ }\textbf {\bibinfo {volume} {365}},\ \bibinfo
  {pages} {367} (\bibinfo {year} {1996})},\ \Eprint
  {http://arxiv.org/abs/hep-ph/9508359} {arXiv:hep-ph/9508359} \BibitemShut
  {NoStop}%
\bibitem [{\citenamefont {Fritzsch}(1997)}]{Fritzsch:1997ps}%
  \BibitemOpen
  \bibfield  {author} {\bibinfo {author} {\bibfnamefont {H.}~\bibnamefont
  {Fritzsch}},\ }\href {\doibase 10.1016/S0370-2693(97)01213-6} {\bibfield
  {journal} {\bibinfo  {journal} {Phys. Lett. B}\ }\textbf {\bibinfo {volume}
  {415}},\ \bibinfo {pages} {83} (\bibinfo {year} {1997})},\ \Eprint
  {http://arxiv.org/abs/hep-ph/9708348} {arXiv:hep-ph/9708348} \BibitemShut
  {NoStop}%
\bibitem [{\citenamefont {Ablikim}\ \emph {et~al.}(2024)\citenamefont {Ablikim}
  \emph {et~al.}}]{BESIII:2023wfi}%
  \BibitemOpen
  \bibfield  {author} {\bibinfo {author} {\bibfnamefont {M.}~\bibnamefont
  {Ablikim}} \emph {et~al.} (\bibinfo {collaboration} {BESIII}),\ }\href
  {\doibase 10.1103/PhysRevLett.132.181901} {\bibfield  {journal} {\bibinfo
  {journal} {Phys. Rev. Lett.}\ }\textbf {\bibinfo {volume} {132}},\ \bibinfo
  {pages} {181901} (\bibinfo {year} {2024})},\ \Eprint
  {http://arxiv.org/abs/2312.05324} {arXiv:2312.05324 [hep-ex]} \BibitemShut
  {NoStop}%
\bibitem [{\citenamefont {Ablikim}\ \emph {et~al.}(2026)\citenamefont {Ablikim}
  \emph {et~al.}}]{BESIII:2026mvn}%
  \BibitemOpen
  \bibfield  {author} {\bibinfo {author} {\bibfnamefont {M.}~\bibnamefont
  {Ablikim}} \emph {et~al.} (\bibinfo {collaboration} {BESIII}),\ }\href@noop
  {} {\  (\bibinfo {year} {2026})},\ \Eprint {http://arxiv.org/abs/2607.20366}
  {arXiv:2607.20366 [hep-ex]} \BibitemShut {NoStop}%
\bibitem [{\citenamefont {Jiang}\ \emph {et~al.}(2023)\citenamefont {Jiang},
  \citenamefont {Sun}, \citenamefont {Chen}, \citenamefont {Chen},
  \citenamefont {Gong}, \citenamefont {Liu},\ and\ \citenamefont
  {Zhang}}]{Jiang:2022ffl}%
  \BibitemOpen
  \bibfield  {author} {\bibinfo {author} {\bibfnamefont {X.}~\bibnamefont
  {Jiang}}, \bibinfo {author} {\bibfnamefont {W.}~\bibnamefont {Sun}}, \bibinfo
  {author} {\bibfnamefont {F.}~\bibnamefont {Chen}}, \bibinfo {author}
  {\bibfnamefont {Y.}~\bibnamefont {Chen}}, \bibinfo {author} {\bibfnamefont
  {M.}~\bibnamefont {Gong}}, \bibinfo {author} {\bibfnamefont {Z.}~\bibnamefont
  {Liu}}, \ and\ \bibinfo {author} {\bibfnamefont {R.}~\bibnamefont {Zhang}},\
  }\href {\doibase 10.1103/PhysRevD.107.094510} {\bibfield  {journal} {\bibinfo
   {journal} {Phys. Rev. D}\ }\textbf {\bibinfo {volume} {107}},\ \bibinfo
  {pages} {094510} (\bibinfo {year} {2023})},\ \Eprint
  {http://arxiv.org/abs/2205.12541} {arXiv:2205.12541 [hep-lat]} \BibitemShut
  {NoStop}%
\bibitem [{\citenamefont {Feldmann}\ \emph {et~al.}(1998)\citenamefont
  {Feldmann}, \citenamefont {Kroll},\ and\ \citenamefont
  {Stech}}]{Feldmann:1998vh}%
  \BibitemOpen
  \bibfield  {author} {\bibinfo {author} {\bibfnamefont {T.}~\bibnamefont
  {Feldmann}}, \bibinfo {author} {\bibfnamefont {P.}~\bibnamefont {Kroll}}, \
  and\ \bibinfo {author} {\bibfnamefont {B.}~\bibnamefont {Stech}},\ }\href
  {\doibase 10.1103/PhysRevD.58.114006} {\bibfield  {journal} {\bibinfo
  {journal} {Phys. Rev. D}\ }\textbf {\bibinfo {volume} {58}},\ \bibinfo
  {pages} {114006} (\bibinfo {year} {1998})},\ \Eprint
  {http://arxiv.org/abs/hep-ph/9802409} {arXiv:hep-ph/9802409} \BibitemShut
  {NoStop}%
\bibitem [{\citenamefont {Feldmann}\ \emph {et~al.}(1999)\citenamefont
  {Feldmann}, \citenamefont {Kroll},\ and\ \citenamefont
  {Stech}}]{Feldmann:1998sh}%
  \BibitemOpen
  \bibfield  {author} {\bibinfo {author} {\bibfnamefont {T.}~\bibnamefont
  {Feldmann}}, \bibinfo {author} {\bibfnamefont {P.}~\bibnamefont {Kroll}}, \
  and\ \bibinfo {author} {\bibfnamefont {B.}~\bibnamefont {Stech}},\ }\href
  {\doibase 10.1016/S0370-2693(99)00085-4} {\bibfield  {journal} {\bibinfo
  {journal} {Phys. Lett. B}\ }\textbf {\bibinfo {volume} {449}},\ \bibinfo
  {pages} {339} (\bibinfo {year} {1999})},\ \Eprint
  {http://arxiv.org/abs/hep-ph/9812269} {arXiv:hep-ph/9812269} \BibitemShut
  {NoStop}%
\bibitem [{\citenamefont {Escribano}\ and\ \citenamefont
  {Frere}(1999)}]{Escribano:1999nh}%
  \BibitemOpen
  \bibfield  {author} {\bibinfo {author} {\bibfnamefont {R.}~\bibnamefont
  {Escribano}}\ and\ \bibinfo {author} {\bibfnamefont {J.~M.}\ \bibnamefont
  {Frere}},\ }\href {\doibase 10.1016/S0370-2693(99)00629-2} {\bibfield
  {journal} {\bibinfo  {journal} {Phys. Lett. B}\ }\textbf {\bibinfo {volume}
  {459}},\ \bibinfo {pages} {288} (\bibinfo {year} {1999})},\ \Eprint
  {http://arxiv.org/abs/hep-ph/9901405} {arXiv:hep-ph/9901405} \BibitemShut
  {NoStop}%
\bibitem [{\citenamefont {Escribano}\ and\ \citenamefont
  {Frere}(2005)}]{Escribano:2005qq}%
  \BibitemOpen
  \bibfield  {author} {\bibinfo {author} {\bibfnamefont {R.}~\bibnamefont
  {Escribano}}\ and\ \bibinfo {author} {\bibfnamefont {J.~M.}\ \bibnamefont
  {Frere}},\ }\href {\doibase 10.1088/1126-6708/2005/06/029} {\bibfield
  {journal} {\bibinfo  {journal} {JHEP}\ }\textbf {\bibinfo {volume} {06}},\
  \bibinfo {pages} {029} (\bibinfo {year} {2005})},\ \Eprint
  {http://arxiv.org/abs/hep-ph/0501072} {arXiv:hep-ph/0501072} \BibitemShut
  {NoStop}%
\bibitem [{\citenamefont {Sakurai}(1960)}]{Sakurai:1960ju}%
  \BibitemOpen
  \bibfield  {author} {\bibinfo {author} {\bibfnamefont {J.~J.}\ \bibnamefont
  {Sakurai}},\ }\href {\doibase 10.1016/0003-4916(60)90126-3} {\bibfield
  {journal} {\bibinfo  {journal} {Annals Phys.}\ }\textbf {\bibinfo {volume}
  {11}},\ \bibinfo {pages} {1} (\bibinfo {year} {1960})}\BibitemShut {NoStop}%
\bibitem [{\citenamefont {Kroll}\ \emph {et~al.}(1967)\citenamefont {Kroll},
  \citenamefont {Lee},\ and\ \citenamefont {Zumino}}]{Kroll:1967it}%
  \BibitemOpen
  \bibfield  {author} {\bibinfo {author} {\bibfnamefont {N.~M.}\ \bibnamefont
  {Kroll}}, \bibinfo {author} {\bibfnamefont {T.~D.}\ \bibnamefont {Lee}}, \
  and\ \bibinfo {author} {\bibfnamefont {B.}~\bibnamefont {Zumino}},\ }\href
  {\doibase 10.1103/PhysRev.157.1376} {\bibfield  {journal} {\bibinfo
  {journal} {Phys. Rev.}\ }\textbf {\bibinfo {volume} {157}},\ \bibinfo {pages}
  {1376} (\bibinfo {year} {1967})}\BibitemShut {NoStop}%
\bibitem [{\citenamefont {Lurie}(1968)}]{Lurie:1968zz}%
  \BibitemOpen
  \bibfield  {author} {\bibinfo {author} {\bibfnamefont {D.}~\bibnamefont
  {Lurie}},\ }\href@noop {} {\emph {\bibinfo {title} {{Particles and
  Fields}}}}\ (\bibinfo  {publisher} {John Wiley and Sons},\ \bibinfo {year}
  {1968})\BibitemShut {NoStop}%
\bibitem [{\citenamefont {Bando}\ \emph {et~al.}(1985)\citenamefont {Bando},
  \citenamefont {Kugo},\ and\ \citenamefont {Yamawaki}}]{Bando:1985rf}%
  \BibitemOpen
  \bibfield  {author} {\bibinfo {author} {\bibfnamefont {M.}~\bibnamefont
  {Bando}}, \bibinfo {author} {\bibfnamefont {T.}~\bibnamefont {Kugo}}, \ and\
  \bibinfo {author} {\bibfnamefont {K.}~\bibnamefont {Yamawaki}},\ }\href
  {\doibase 10.1016/0550-3213(85)90647-9} {\bibfield  {journal} {\bibinfo
  {journal} {Nucl. Phys. B}\ }\textbf {\bibinfo {volume} {259}},\ \bibinfo
  {pages} {493} (\bibinfo {year} {1985})}\BibitemShut {NoStop}%
\bibitem [{\citenamefont {Bando}\ \emph {et~al.}(1988)\citenamefont {Bando},
  \citenamefont {Kugo},\ and\ \citenamefont {Yamawaki}}]{Bando:1987br}%
  \BibitemOpen
  \bibfield  {author} {\bibinfo {author} {\bibfnamefont {M.}~\bibnamefont
  {Bando}}, \bibinfo {author} {\bibfnamefont {T.}~\bibnamefont {Kugo}}, \ and\
  \bibinfo {author} {\bibfnamefont {K.}~\bibnamefont {Yamawaki}},\ }\href
  {\doibase 10.1016/0370-1573(88)90019-1} {\bibfield  {journal} {\bibinfo
  {journal} {Phys. Rept.}\ }\textbf {\bibinfo {volume} {164}},\ \bibinfo
  {pages} {217} (\bibinfo {year} {1988})}\BibitemShut {NoStop}%
\bibitem [{\citenamefont {Meissner}(1988)}]{Meissner:1987ge}%
  \BibitemOpen
  \bibfield  {author} {\bibinfo {author} {\bibfnamefont {U.~G.}\ \bibnamefont
  {Meissner}},\ }\href {\doibase 10.1016/0370-1573(88)90090-7} {\bibfield
  {journal} {\bibinfo  {journal} {Phys. Rept.}\ }\textbf {\bibinfo {volume}
  {161}},\ \bibinfo {pages} {213} (\bibinfo {year} {1988})}\BibitemShut
  {NoStop}%
\bibitem [{\citenamefont {Bramon}\ \emph {et~al.}(2001)\citenamefont {Bramon},
  \citenamefont {Escribano},\ and\ \citenamefont {Scadron}}]{Bramon:2000fr}%
  \BibitemOpen
  \bibfield  {author} {\bibinfo {author} {\bibfnamefont {A.}~\bibnamefont
  {Bramon}}, \bibinfo {author} {\bibfnamefont {R.}~\bibnamefont {Escribano}}, \
  and\ \bibinfo {author} {\bibfnamefont {M.~D.}\ \bibnamefont {Scadron}},\
  }\href {\doibase 10.1016/S0370-2693(01)00161-7} {\bibfield  {journal}
  {\bibinfo  {journal} {Phys. Lett. B}\ }\textbf {\bibinfo {volume} {503}},\
  \bibinfo {pages} {271} (\bibinfo {year} {2001})},\ \Eprint
  {http://arxiv.org/abs/hep-ph/0012049} {arXiv:hep-ph/0012049} \BibitemShut
  {NoStop}%
\bibitem [{\citenamefont {Kou}(2001)}]{Kou:1999tt}%
  \BibitemOpen
  \bibfield  {author} {\bibinfo {author} {\bibfnamefont {E.}~\bibnamefont
  {Kou}},\ }\href {\doibase 10.1103/PhysRevD.63.054027} {\bibfield  {journal}
  {\bibinfo  {journal} {Phys. Rev. D}\ }\textbf {\bibinfo {volume} {63}},\
  \bibinfo {pages} {054027} (\bibinfo {year} {2001})},\ \Eprint
  {http://arxiv.org/abs/hep-ph/9908214} {arXiv:hep-ph/9908214} \BibitemShut
  {NoStop}%
\bibitem [{\citenamefont {Ambrosino}\ \emph {et~al.}(2007)\citenamefont
  {Ambrosino} \emph {et~al.}}]{KLOE:2006guu}%
  \BibitemOpen
  \bibfield  {author} {\bibinfo {author} {\bibfnamefont {F.}~\bibnamefont
  {Ambrosino}} \emph {et~al.} (\bibinfo {collaboration} {KLOE}),\ }\href
  {\doibase 10.1016/j.physletb.2007.03.032} {\bibfield  {journal} {\bibinfo
  {journal} {Phys. Lett. B}\ }\textbf {\bibinfo {volume} {648}},\ \bibinfo
  {pages} {267} (\bibinfo {year} {2007})},\ \Eprint
  {http://arxiv.org/abs/hep-ex/0612029} {arXiv:hep-ex/0612029} \BibitemShut
  {NoStop}%
\bibitem [{\citenamefont {Aloisio}\ \emph {et~al.}(2002)\citenamefont {Aloisio}
  \emph {et~al.}}]{KLOE:2002jed}%
  \BibitemOpen
  \bibfield  {author} {\bibinfo {author} {\bibfnamefont {A.}~\bibnamefont
  {Aloisio}} \emph {et~al.} (\bibinfo {collaboration} {KLOE}),\ }\href
  {\doibase 10.1016/S0370-2693(02)02145-7} {\bibfield  {journal} {\bibinfo
  {journal} {Phys. Lett. B}\ }\textbf {\bibinfo {volume} {541}},\ \bibinfo
  {pages} {45} (\bibinfo {year} {2002})},\ \Eprint
  {http://arxiv.org/abs/hep-ex/0206010} {arXiv:hep-ex/0206010} \BibitemShut
  {NoStop}%
\bibitem [{\citenamefont {Ambrosino}\ \emph {et~al.}(2009)\citenamefont
  {Ambrosino} \emph {et~al.}}]{Ambrosino:2009sc}%
  \BibitemOpen
  \bibfield  {author} {\bibinfo {author} {\bibfnamefont {F.}~\bibnamefont
  {Ambrosino}} \emph {et~al.},\ }\href {\doibase 10.1088/1126-6708/2009/07/105}
  {\bibfield  {journal} {\bibinfo  {journal} {JHEP}\ }\textbf {\bibinfo
  {volume} {07}},\ \bibinfo {pages} {105} (\bibinfo {year} {2009})},\ \Eprint
  {http://arxiv.org/abs/0906.3819} {arXiv:0906.3819 [hep-ph]} \BibitemShut
  {NoStop}%
\bibitem [{\citenamefont {Escribano}\ and\ \citenamefont
  {Nadal}(2007)}]{Escribano:2007cd}%
  \BibitemOpen
  \bibfield  {author} {\bibinfo {author} {\bibfnamefont {R.}~\bibnamefont
  {Escribano}}\ and\ \bibinfo {author} {\bibfnamefont {J.}~\bibnamefont
  {Nadal}},\ }\href {\doibase 10.1088/1126-6708/2007/05/006} {\bibfield
  {journal} {\bibinfo  {journal} {JHEP}\ }\textbf {\bibinfo {volume} {05}},\
  \bibinfo {pages} {006} (\bibinfo {year} {2007})},\ \Eprint
  {http://arxiv.org/abs/hep-ph/0703187} {arXiv:hep-ph/0703187} \BibitemShut
  {NoStop}%
\bibitem [{\citenamefont {Thomas}(2007)}]{Thomas:2007uy}%
  \BibitemOpen
  \bibfield  {author} {\bibinfo {author} {\bibfnamefont {C.~E.}\ \bibnamefont
  {Thomas}},\ }\href {\doibase 10.1088/1126-6708/2007/10/026} {\bibfield
  {journal} {\bibinfo  {journal} {JHEP}\ }\textbf {\bibinfo {volume} {10}},\
  \bibinfo {pages} {026} (\bibinfo {year} {2007})},\ \Eprint
  {http://arxiv.org/abs/0705.1500} {arXiv:0705.1500 [hep-ph]} \BibitemShut
  {NoStop}%
\bibitem [{\citenamefont {Takahashi}\ \emph {et~al.}(2026)\citenamefont
  {Takahashi} \emph {et~al.}}]{ParticleDataGroup:2026aaa}%
  \BibitemOpen
  \bibfield  {author} {\bibinfo {author} {\bibfnamefont {F.}~\bibnamefont
  {Takahashi}} \emph {et~al.} (\bibinfo {collaboration} {Particle Data
  Group}),\ }\href {\doibase 10.1142/S0217751X26300115} {\bibfield  {journal}
  {\bibinfo  {journal} {Int. J. Mod. Phys. A}\ }\textbf {\bibinfo {volume}
  {41}},\ \bibinfo {pages} {2630011} (\bibinfo {year} {2026})}\BibitemShut
  {NoStop}%
\bibitem [{\citenamefont {Escribano}\ and\ \citenamefont
  {Royo}(2020)}]{Escribano:2020jdy}%
  \BibitemOpen
  \bibfield  {author} {\bibinfo {author} {\bibfnamefont {R.}~\bibnamefont
  {Escribano}}\ and\ \bibinfo {author} {\bibfnamefont {E.}~\bibnamefont
  {Royo}},\ }\href {\doibase 10.1016/j.physletb.2020.135534} {\bibfield
  {journal} {\bibinfo  {journal} {Phys. Lett. B}\ }\textbf {\bibinfo {volume}
  {807}},\ \bibinfo {pages} {135534} (\bibinfo {year} {2020})},\ \Eprint
  {http://arxiv.org/abs/2003.08379} {arXiv:2003.08379 [hep-ph]} \BibitemShut
  {NoStop}%
\bibitem [{\citenamefont {Cheng}\ \emph {et~al.}(2009)\citenamefont {Cheng},
  \citenamefont {Li},\ and\ \citenamefont {Liu}}]{Cheng:2008ss}%
  \BibitemOpen
  \bibfield  {author} {\bibinfo {author} {\bibfnamefont {H.-Y.}\ \bibnamefont
  {Cheng}}, \bibinfo {author} {\bibfnamefont {H.-n.}\ \bibnamefont {Li}}, \
  and\ \bibinfo {author} {\bibfnamefont {K.-F.}\ \bibnamefont {Liu}},\ }\href
  {\doibase 10.1103/PhysRevD.79.014024} {\bibfield  {journal} {\bibinfo
  {journal} {Phys. Rev. D}\ }\textbf {\bibinfo {volume} {79}},\ \bibinfo
  {pages} {014024} (\bibinfo {year} {2009})},\ \Eprint
  {http://arxiv.org/abs/0811.2577} {arXiv:0811.2577 [hep-ph]} \BibitemShut
  {NoStop}%
\bibitem [{\citenamefont {Mathieu}\ and\ \citenamefont
  {Vento}(2010{\natexlab{a}})}]{Mathieu:2009sg}%
  \BibitemOpen
  \bibfield  {author} {\bibinfo {author} {\bibfnamefont {V.}~\bibnamefont
  {Mathieu}}\ and\ \bibinfo {author} {\bibfnamefont {V.}~\bibnamefont
  {Vento}},\ }\href {\doibase 10.1103/PhysRevD.81.034004} {\bibfield  {journal}
  {\bibinfo  {journal} {Phys. Rev. D}\ }\textbf {\bibinfo {volume} {81}},\
  \bibinfo {pages} {034004} (\bibinfo {year} {2010}{\natexlab{a}})},\ \Eprint
  {http://arxiv.org/abs/0910.0212} {arXiv:0910.0212 [hep-ph]} \BibitemShut
  {NoStop}%
\bibitem [{\citenamefont {Cohen}\ and\ \citenamefont
  {Lipkin}(1979)}]{Cohen:1978ge}%
  \BibitemOpen
  \bibfield  {author} {\bibinfo {author} {\bibfnamefont {I.}~\bibnamefont
  {Cohen}}\ and\ \bibinfo {author} {\bibfnamefont {H.~J.}\ \bibnamefont
  {Lipkin}},\ }\href {\doibase 10.1016/0550-3213(79)90424-3} {\bibfield
  {journal} {\bibinfo  {journal} {Nucl. Phys. B}\ }\textbf {\bibinfo {volume}
  {151}},\ \bibinfo {pages} {16} (\bibinfo {year} {1979})}\BibitemShut
  {NoStop}%
\bibitem [{\citenamefont {Klempt}\ and\ \citenamefont
  {Zaitsev}(2007)}]{Klempt:2007cp}%
  \BibitemOpen
  \bibfield  {author} {\bibinfo {author} {\bibfnamefont {E.}~\bibnamefont
  {Klempt}}\ and\ \bibinfo {author} {\bibfnamefont {A.}~\bibnamefont
  {Zaitsev}},\ }\href {\doibase 10.1016/j.physrep.2007.07.006} {\bibfield
  {journal} {\bibinfo  {journal} {Phys. Rept.}\ }\textbf {\bibinfo {volume}
  {454}},\ \bibinfo {pages} {1} (\bibinfo {year} {2007})},\ \Eprint
  {http://arxiv.org/abs/0708.4016} {arXiv:0708.4016 [hep-ph]} \BibitemShut
  {NoStop}%
\bibitem [{\citenamefont {Damgaard}\ \emph {et~al.}(1994)\citenamefont
  {Damgaard}, \citenamefont {Nielsen},\ and\ \citenamefont
  {Sollacher}}]{Damgaard:1993sx}%
  \BibitemOpen
  \bibfield  {author} {\bibinfo {author} {\bibfnamefont {P.~H.}\ \bibnamefont
  {Damgaard}}, \bibinfo {author} {\bibfnamefont {H.~B.}\ \bibnamefont
  {Nielsen}}, \ and\ \bibinfo {author} {\bibfnamefont {R.}~\bibnamefont
  {Sollacher}},\ }\href {\doibase 10.1016/0550-3213(94)90254-2} {\bibfield
  {journal} {\bibinfo  {journal} {Nucl. Phys. B}\ }\textbf {\bibinfo {volume}
  {414}},\ \bibinfo {pages} {541} (\bibinfo {year} {1994})},\ \Eprint
  {http://arxiv.org/abs/hep-ph/9308259} {arXiv:hep-ph/9308259} \BibitemShut
  {NoStop}%
\bibitem [{\citenamefont {Bramon}\ \emph {et~al.}(1999)\citenamefont {Bramon},
  \citenamefont {Escribano},\ and\ \citenamefont {Scadron}}]{Bramon:1997va}%
  \BibitemOpen
  \bibfield  {author} {\bibinfo {author} {\bibfnamefont {A.}~\bibnamefont
  {Bramon}}, \bibinfo {author} {\bibfnamefont {R.}~\bibnamefont {Escribano}}, \
  and\ \bibinfo {author} {\bibfnamefont {M.~D.}\ \bibnamefont {Scadron}},\
  }\href {\doibase 10.1007/s100529801009} {\bibfield  {journal} {\bibinfo
  {journal} {Eur. Phys. J. C}\ }\textbf {\bibinfo {volume} {7}},\ \bibinfo
  {pages} {271} (\bibinfo {year} {1999})},\ \Eprint
  {http://arxiv.org/abs/hep-ph/9711229} {arXiv:hep-ph/9711229} \BibitemShut
  {NoStop}%
\bibitem [{\citenamefont {Mathieu}\ and\ \citenamefont
  {Vento}(2010{\natexlab{b}})}]{Mathieu:2010ss}%
  \BibitemOpen
  \bibfield  {author} {\bibinfo {author} {\bibfnamefont {V.}~\bibnamefont
  {Mathieu}}\ and\ \bibinfo {author} {\bibfnamefont {V.}~\bibnamefont
  {Vento}},\ }\href {\doibase 10.1016/j.physletb.2010.04.021} {\bibfield
  {journal} {\bibinfo  {journal} {Phys. Lett. B}\ }\textbf {\bibinfo {volume}
  {688}},\ \bibinfo {pages} {314} (\bibinfo {year} {2010}{\natexlab{b}})},\
  \Eprint {http://arxiv.org/abs/1003.2119} {arXiv:1003.2119 [hep-ph]}
  \BibitemShut {NoStop}%
\bibitem [{\citenamefont {Ottnad}\ \emph {et~al.}(2025)\citenamefont {Ottnad},
  \citenamefont {Bacchio}, \citenamefont {Finkenrath}, \citenamefont
  {Kostrzewa}, \citenamefont {Petschlies}, \citenamefont {Pittler},
  \citenamefont {Urbach},\ and\ \citenamefont {Wenger}}]{Ottnad:2025zxq}%
  \BibitemOpen
  \bibfield  {author} {\bibinfo {author} {\bibfnamefont {K.}~\bibnamefont
  {Ottnad}}, \bibinfo {author} {\bibfnamefont {S.}~\bibnamefont {Bacchio}},
  \bibinfo {author} {\bibfnamefont {J.}~\bibnamefont {Finkenrath}}, \bibinfo
  {author} {\bibfnamefont {B.}~\bibnamefont {Kostrzewa}}, \bibinfo {author}
  {\bibfnamefont {M.}~\bibnamefont {Petschlies}}, \bibinfo {author}
  {\bibfnamefont {F.}~\bibnamefont {Pittler}}, \bibinfo {author} {\bibfnamefont
  {C.}~\bibnamefont {Urbach}}, \ and\ \bibinfo {author} {\bibfnamefont
  {U.}~\bibnamefont {Wenger}},\ }\href {\doibase
  10.1140/epja/s10050-025-01635-0} {\bibfield  {journal} {\bibinfo  {journal}
  {Eur. Phys. J. A}\ }\textbf {\bibinfo {volume} {61}},\ \bibinfo {pages} {169}
  (\bibinfo {year} {2025})},\ \bibinfo {note} {[Erratum: Eur.Phys.J.A 61, 227
  (2025)]},\ \Eprint {http://arxiv.org/abs/2503.09895} {arXiv:2503.09895
  [hep-lat]} \BibitemShut {NoStop}%
\bibitem [{\citenamefont {Chen}\ \emph {et~al.}(2006)\citenamefont {Chen} \emph
  {et~al.}}]{Chen:2005mg}%
  \BibitemOpen
  \bibfield  {author} {\bibinfo {author} {\bibfnamefont {Y.}~\bibnamefont
  {Chen}} \emph {et~al.},\ }\href {\doibase 10.1103/PhysRevD.73.014516}
  {\bibfield  {journal} {\bibinfo  {journal} {Phys. Rev. D}\ }\textbf {\bibinfo
  {volume} {73}},\ \bibinfo {pages} {014516} (\bibinfo {year} {2006})},\
  \Eprint {http://arxiv.org/abs/hep-lat/0510074} {arXiv:hep-lat/0510074}
  \BibitemShut {NoStop}%
\bibitem [{\citenamefont {Gregory}\ \emph {et~al.}(2012)\citenamefont
  {Gregory}, \citenamefont {Irving}, \citenamefont {Lucini}, \citenamefont
  {McNeile}, \citenamefont {Rago}, \citenamefont {Richards},\ and\
  \citenamefont {Rinaldi}}]{Gregory:2012hu}%
  \BibitemOpen
  \bibfield  {author} {\bibinfo {author} {\bibfnamefont {E.}~\bibnamefont
  {Gregory}}, \bibinfo {author} {\bibfnamefont {A.}~\bibnamefont {Irving}},
  \bibinfo {author} {\bibfnamefont {B.}~\bibnamefont {Lucini}}, \bibinfo
  {author} {\bibfnamefont {C.}~\bibnamefont {McNeile}}, \bibinfo {author}
  {\bibfnamefont {A.}~\bibnamefont {Rago}}, \bibinfo {author} {\bibfnamefont
  {C.}~\bibnamefont {Richards}}, \ and\ \bibinfo {author} {\bibfnamefont
  {E.}~\bibnamefont {Rinaldi}},\ }\href {\doibase 10.1007/JHEP10(2012)170}
  {\bibfield  {journal} {\bibinfo  {journal} {JHEP}\ }\textbf {\bibinfo
  {volume} {10}},\ \bibinfo {pages} {170} (\bibinfo {year} {2012})},\ \Eprint
  {http://arxiv.org/abs/1208.1858} {arXiv:1208.1858 [hep-lat]} \BibitemShut
  {NoStop}%
\bibitem [{\citenamefont {Athenodorou}\ and\ \citenamefont
  {Teper}(2020)}]{Athenodorou:2020ani}%
  \BibitemOpen
  \bibfield  {author} {\bibinfo {author} {\bibfnamefont {A.}~\bibnamefont
  {Athenodorou}}\ and\ \bibinfo {author} {\bibfnamefont {M.}~\bibnamefont
  {Teper}},\ }\href {\doibase 10.1007/JHEP11(2020)172} {\bibfield  {journal}
  {\bibinfo  {journal} {JHEP}\ }\textbf {\bibinfo {volume} {11}},\ \bibinfo
  {pages} {172} (\bibinfo {year} {2020})},\ \Eprint
  {http://arxiv.org/abs/2007.06422} {arXiv:2007.06422 [hep-lat]} \BibitemShut
  {NoStop}%
\bibitem [{\citenamefont {Huang}\ \emph {et~al.}(2025)\citenamefont {Huang},
  \citenamefont {Jin},\ and\ \citenamefont {Zhang}}]{Huang:2025pyv}%
  \BibitemOpen
  \bibfield  {author} {\bibinfo {author} {\bibfnamefont {Y.}~\bibnamefont
  {Huang}}, \bibinfo {author} {\bibfnamefont {S.}~\bibnamefont {Jin}}, \ and\
  \bibinfo {author} {\bibfnamefont {P.}~\bibnamefont {Zhang}},\ }\href
  {\doibase 10.1142/S0217751X25300078} {\bibfield  {journal} {\bibinfo
  {journal} {Int. J. Mod. Phys. A}\ }\textbf {\bibinfo {volume} {40}},\
  \bibinfo {pages} {2530007} (\bibinfo {year} {2025})},\ \Eprint
  {http://arxiv.org/abs/2503.13286} {arXiv:2503.13286 [hep-ex]} \BibitemShut
  {NoStop}%
\bibitem [{\citenamefont {Zhang}\ \emph {et~al.}(2026)\citenamefont {Zhang},
  \citenamefont {Wan},\ and\ \citenamefont {Qiao}}]{Zhang:2026poz}%
  \BibitemOpen
  \bibfield  {author} {\bibinfo {author} {\bibfnamefont {S.-Q.}\ \bibnamefont
  {Zhang}}, \bibinfo {author} {\bibfnamefont {B.-D.}\ \bibnamefont {Wan}}, \
  and\ \bibinfo {author} {\bibfnamefont {C.-F.}\ \bibnamefont {Qiao}},\
  }\href@noop {} {\  (\bibinfo {year} {2026})},\ \Eprint
  {http://arxiv.org/abs/2609.05908} {arXiv:2609.05908 [hep-ph]} \BibitemShut
  {NoStop}%
\bibitem [{\citenamefont {Godfrey}\ and\ \citenamefont
  {Isgur}(1985)}]{Godfrey:1985xj}%
  \BibitemOpen
  \bibfield  {author} {\bibinfo {author} {\bibfnamefont {S.}~\bibnamefont
  {Godfrey}}\ and\ \bibinfo {author} {\bibfnamefont {N.}~\bibnamefont
  {Isgur}},\ }\href {\doibase 10.1103/PhysRevD.32.189} {\bibfield  {journal}
  {\bibinfo  {journal} {Phys. Rev. D}\ }\textbf {\bibinfo {volume} {32}},\
  \bibinfo {pages} {189} (\bibinfo {year} {1985})}\BibitemShut {NoStop}%
\bibitem [{\citenamefont {Cheng}\ and\ \citenamefont
  {Zhao}(2022)}]{Cheng:2021nal}%
  \BibitemOpen
  \bibfield  {author} {\bibinfo {author} {\bibfnamefont {Y.}~\bibnamefont
  {Cheng}}\ and\ \bibinfo {author} {\bibfnamefont {Q.}~\bibnamefont {Zhao}},\
  }\href {\doibase 10.1103/PhysRevD.105.076023} {\bibfield  {journal} {\bibinfo
   {journal} {Phys. Rev. D}\ }\textbf {\bibinfo {volume} {105}},\ \bibinfo
  {pages} {076023} (\bibinfo {year} {2022})},\ \Eprint
  {http://arxiv.org/abs/2106.12483} {arXiv:2106.12483 [hep-ph]} \BibitemShut
  {NoStop}%
\bibitem [{\citenamefont {Peccei}\ and\ \citenamefont
  {Quinn}(1977{\natexlab{a}})}]{Peccei:1977hh}%
  \BibitemOpen
  \bibfield  {author} {\bibinfo {author} {\bibfnamefont {R.~D.}\ \bibnamefont
  {Peccei}}\ and\ \bibinfo {author} {\bibfnamefont {H.~R.}\ \bibnamefont
  {Quinn}},\ }\href {\doibase 10.1103/PhysRevLett.38.1440} {\bibfield
  {journal} {\bibinfo  {journal} {Phys. Rev. Lett.}\ }\textbf {\bibinfo
  {volume} {38}},\ \bibinfo {pages} {1440} (\bibinfo {year}
  {1977}{\natexlab{a}})}\BibitemShut {NoStop}%
\bibitem [{\citenamefont {Peccei}\ and\ \citenamefont
  {Quinn}(1977{\natexlab{b}})}]{Peccei:1977ur}%
  \BibitemOpen
  \bibfield  {author} {\bibinfo {author} {\bibfnamefont {R.~D.}\ \bibnamefont
  {Peccei}}\ and\ \bibinfo {author} {\bibfnamefont {H.~R.}\ \bibnamefont
  {Quinn}},\ }\href {\doibase 10.1103/PhysRevD.16.1791} {\bibfield  {journal}
  {\bibinfo  {journal} {Phys. Rev. D}\ }\textbf {\bibinfo {volume} {16}},\
  \bibinfo {pages} {1791} (\bibinfo {year} {1977}{\natexlab{b}})}\BibitemShut
  {NoStop}%
\bibitem [{\citenamefont {Weinberg}(1978)}]{Weinberg:1977ma}%
  \BibitemOpen
  \bibfield  {author} {\bibinfo {author} {\bibfnamefont {S.}~\bibnamefont
  {Weinberg}},\ }\href {\doibase 10.1103/PhysRevLett.40.223} {\bibfield
  {journal} {\bibinfo  {journal} {Phys. Rev. Lett.}\ }\textbf {\bibinfo
  {volume} {40}},\ \bibinfo {pages} {223} (\bibinfo {year} {1978})}\BibitemShut
  {NoStop}%
\bibitem [{\citenamefont {Wilczek}(1978)}]{Wilczek:1977pj}%
  \BibitemOpen
  \bibfield  {author} {\bibinfo {author} {\bibfnamefont {F.}~\bibnamefont
  {Wilczek}},\ }\href {\doibase 10.1103/PhysRevLett.40.279} {\bibfield
  {journal} {\bibinfo  {journal} {Phys. Rev. Lett.}\ }\textbf {\bibinfo
  {volume} {40}},\ \bibinfo {pages} {279} (\bibinfo {year} {1978})}\BibitemShut
  {NoStop}%
\bibitem [{\citenamefont {Kim}(1987)}]{Kim:1986ax}%
  \BibitemOpen
  \bibfield  {author} {\bibinfo {author} {\bibfnamefont {J.~E.}\ \bibnamefont
  {Kim}},\ }\href {\doibase 10.1016/0370-1573(87)90017-2} {\bibfield  {journal}
  {\bibinfo  {journal} {Phys. Rept.}\ }\textbf {\bibinfo {volume} {150}},\
  \bibinfo {pages} {1} (\bibinfo {year} {1987})}\BibitemShut {NoStop}%
\bibitem [{\citenamefont {Cheng}(1988)}]{Cheng:1987gp}%
  \BibitemOpen
  \bibfield  {author} {\bibinfo {author} {\bibfnamefont {H.-Y.}\ \bibnamefont
  {Cheng}},\ }\href {\doibase 10.1016/0370-1573(88)90135-4} {\bibfield
  {journal} {\bibinfo  {journal} {Phys. Rept.}\ }\textbf {\bibinfo {volume}
  {158}},\ \bibinfo {pages} {1} (\bibinfo {year} {1988})}\BibitemShut {NoStop}%
\bibitem [{\citenamefont {Kim}\ and\ \citenamefont
  {Carosi}(2010)}]{Kim:2008hd}%
  \BibitemOpen
  \bibfield  {author} {\bibinfo {author} {\bibfnamefont {J.~E.}\ \bibnamefont
  {Kim}}\ and\ \bibinfo {author} {\bibfnamefont {G.}~\bibnamefont {Carosi}},\
  }\href {\doibase 10.1103/RevModPhys.82.557} {\bibfield  {journal} {\bibinfo
  {journal} {Rev. Mod. Phys.}\ }\textbf {\bibinfo {volume} {82}},\ \bibinfo
  {pages} {557} (\bibinfo {year} {2010})},\ \bibinfo {note} {[Erratum:
  Rev.Mod.Phys. 91, 049902 (2019)]},\ \Eprint {http://arxiv.org/abs/0807.3125}
  {arXiv:0807.3125 [hep-ph]} \BibitemShut {NoStop}%
\bibitem [{\citenamefont {Marsh}(2016)}]{Marsh:2015xka}%
  \BibitemOpen
  \bibfield  {author} {\bibinfo {author} {\bibfnamefont {D.~J.~E.}\
  \bibnamefont {Marsh}},\ }\href {\doibase 10.1016/j.physrep.2016.06.005}
  {\bibfield  {journal} {\bibinfo  {journal} {Phys. Rept.}\ }\textbf {\bibinfo
  {volume} {643}},\ \bibinfo {pages} {1} (\bibinfo {year} {2016})},\ \Eprint
  {http://arxiv.org/abs/1510.07633} {arXiv:1510.07633 [astro-ph.CO]}
  \BibitemShut {NoStop}%
\bibitem [{\citenamefont {Alves}\ and\ \citenamefont
  {Gonz{\`a}lez-Sol{\'\i}s}(2024)}]{Alves:2024dpa}%
  \BibitemOpen
  \bibfield  {author} {\bibinfo {author} {\bibfnamefont {D.~S.~M.}\
  \bibnamefont {Alves}}\ and\ \bibinfo {author} {\bibfnamefont
  {S.}~\bibnamefont {Gonz{\`a}lez-Sol{\'\i}s}},\ }\href {\doibase
  10.1007/JHEP07(2024)264} {\bibfield  {journal} {\bibinfo  {journal} {JHEP}\
  }\textbf {\bibinfo {volume} {07}},\ \bibinfo {pages} {264} (\bibinfo {year}
  {2024})},\ \Eprint {http://arxiv.org/abs/2402.02993} {arXiv:2402.02993
  [hep-ph]} \BibitemShut {NoStop}%
\bibitem [{\citenamefont {Depta}\ \emph {et~al.}(2020)\citenamefont {Depta},
  \citenamefont {Hufnagel},\ and\ \citenamefont
  {Schmidt-Hoberg}}]{Depta:2020wmr}%
  \BibitemOpen
  \bibfield  {author} {\bibinfo {author} {\bibfnamefont {P.~F.}\ \bibnamefont
  {Depta}}, \bibinfo {author} {\bibfnamefont {M.}~\bibnamefont {Hufnagel}}, \
  and\ \bibinfo {author} {\bibfnamefont {K.}~\bibnamefont {Schmidt-Hoberg}},\
  }\href {\doibase 10.1088/1475-7516/2020/05/009} {\bibfield  {journal}
  {\bibinfo  {journal} {JCAP}\ }\textbf {\bibinfo {volume} {05}},\ \bibinfo
  {pages} {009} (\bibinfo {year} {2020})},\ \Eprint
  {http://arxiv.org/abs/2002.08370} {arXiv:2002.08370 [hep-ph]} \BibitemShut
  {NoStop}%
\bibitem [{\citenamefont {Bauer}\ \emph {et~al.}(2022)\citenamefont {Bauer},
  \citenamefont {Neubert}, \citenamefont {Renner}, \citenamefont {Schnubel},\
  and\ \citenamefont {Thamm}}]{Bauer:2021mvw}%
  \BibitemOpen
  \bibfield  {author} {\bibinfo {author} {\bibfnamefont {M.}~\bibnamefont
  {Bauer}}, \bibinfo {author} {\bibfnamefont {M.}~\bibnamefont {Neubert}},
  \bibinfo {author} {\bibfnamefont {S.}~\bibnamefont {Renner}}, \bibinfo
  {author} {\bibfnamefont {M.}~\bibnamefont {Schnubel}}, \ and\ \bibinfo
  {author} {\bibfnamefont {A.}~\bibnamefont {Thamm}},\ }\href {\doibase
  10.1007/JHEP09(2022)056} {\bibfield  {journal} {\bibinfo  {journal} {JHEP}\
  }\textbf {\bibinfo {volume} {09}},\ \bibinfo {pages} {056} (\bibinfo {year}
  {2022})},\ \Eprint {http://arxiv.org/abs/2110.10698} {arXiv:2110.10698
  [hep-ph]} \BibitemShut {NoStop}%
\bibitem [{\citenamefont {Alves}\ and\ \citenamefont
  {Weiner}(2018)}]{Alves:2017avw}%
  \BibitemOpen
  \bibfield  {author} {\bibinfo {author} {\bibfnamefont {D.~S.~M.}\
  \bibnamefont {Alves}}\ and\ \bibinfo {author} {\bibfnamefont
  {N.}~\bibnamefont {Weiner}},\ }\href {\doibase 10.1007/JHEP07(2018)092}
  {\bibfield  {journal} {\bibinfo  {journal} {JHEP}\ }\textbf {\bibinfo
  {volume} {07}},\ \bibinfo {pages} {092} (\bibinfo {year} {2018})},\ \Eprint
  {http://arxiv.org/abs/1710.03764} {arXiv:1710.03764 [hep-ph]} \BibitemShut
  {NoStop}%
\bibitem [{\citenamefont {Cornella}\ \emph {et~al.}(2020)\citenamefont
  {Cornella}, \citenamefont {Paradisi},\ and\ \citenamefont
  {Sumensari}}]{Cornella:2019uxs}%
  \BibitemOpen
  \bibfield  {author} {\bibinfo {author} {\bibfnamefont {C.}~\bibnamefont
  {Cornella}}, \bibinfo {author} {\bibfnamefont {P.}~\bibnamefont {Paradisi}},
  \ and\ \bibinfo {author} {\bibfnamefont {O.}~\bibnamefont {Sumensari}},\
  }\href {\doibase 10.1007/JHEP01(2020)158} {\bibfield  {journal} {\bibinfo
  {journal} {JHEP}\ }\textbf {\bibinfo {volume} {01}},\ \bibinfo {pages} {158}
  (\bibinfo {year} {2020})},\ \Eprint {http://arxiv.org/abs/1911.06279}
  {arXiv:1911.06279 [hep-ph]} \BibitemShut {NoStop}%
\bibitem [{\citenamefont {Chang}\ \emph {et~al.}(2018)\citenamefont {Chang},
  \citenamefont {Essig},\ and\ \citenamefont {McDermott}}]{Chang:2018rso}%
  \BibitemOpen
  \bibfield  {author} {\bibinfo {author} {\bibfnamefont {J.~H.}\ \bibnamefont
  {Chang}}, \bibinfo {author} {\bibfnamefont {R.}~\bibnamefont {Essig}}, \ and\
  \bibinfo {author} {\bibfnamefont {S.~D.}\ \bibnamefont {McDermott}},\ }\href
  {\doibase 10.1007/JHEP09(2018)051} {\bibfield  {journal} {\bibinfo  {journal}
  {JHEP}\ }\textbf {\bibinfo {volume} {09}},\ \bibinfo {pages} {051} (\bibinfo
  {year} {2018})},\ \Eprint {http://arxiv.org/abs/1803.00993} {arXiv:1803.00993
  [hep-ph]} \BibitemShut {NoStop}%
\bibitem [{\citenamefont {Raffelt}(2008)}]{Raffelt:2006cw}%
  \BibitemOpen
  \bibfield  {author} {\bibinfo {author} {\bibfnamefont {G.~G.}\ \bibnamefont
  {Raffelt}},\ }\href {\doibase 10.1007/978-3-540-73518-2_3} {\bibfield
  {journal} {\bibinfo  {journal} {Lect. Notes Phys.}\ }\textbf {\bibinfo
  {volume} {741}},\ \bibinfo {pages} {51} (\bibinfo {year} {2008})},\ \Eprint
  {http://arxiv.org/abs/hep-ph/0611350} {arXiv:hep-ph/0611350} \BibitemShut
  {NoStop}%
\bibitem [{\citenamefont {Bauer}\ \emph {et~al.}(2017)\citenamefont {Bauer},
  \citenamefont {Neubert},\ and\ \citenamefont {Thamm}}]{Bauer:2017ris}%
  \BibitemOpen
  \bibfield  {author} {\bibinfo {author} {\bibfnamefont {M.}~\bibnamefont
  {Bauer}}, \bibinfo {author} {\bibfnamefont {M.}~\bibnamefont {Neubert}}, \
  and\ \bibinfo {author} {\bibfnamefont {A.}~\bibnamefont {Thamm}},\ }\href
  {\doibase 10.1007/JHEP12(2017)044} {\bibfield  {journal} {\bibinfo  {journal}
  {JHEP}\ }\textbf {\bibinfo {volume} {12}},\ \bibinfo {pages} {044} (\bibinfo
  {year} {2017})},\ \Eprint {http://arxiv.org/abs/1708.00443} {arXiv:1708.00443
  [hep-ph]} \BibitemShut {NoStop}%
\end{thebibliography}%

\end{document}